\pdfoutput=1
\documentclass[11pt,twoside,a4paper,cmspaper,final,collab]{cms-tdr}
\def\svnVersion{c1df3bd}\def\svnDate{2023/09/20}\def\cmsCernNoTag{CERN-EP-2023-003}\def\cmsCernDate{\today}\def\cmsMessage{Published in the European Physical Journal C as \href{http://dx.doi.org/10.1140/epjc/s10052-023-11687-5}{\doi{10.1140/epjc/s10052-023-11687-5}.}}
\begin{document}\cmsNoteHeader{EXO-19-009}

\newlength\cmsTabSkip\setlength{\cmsTabSkip}{1ex}
\newcommand{\mmiss}{\ensuremath{m_{\text{miss}}}\xspace}
\newcommand{\mX}{\ensuremath{m_{\PX}}\xspace}
\renewcommand{\cmsCollabName}{The CMS and TOTEM Collaborations}
\renewcommand{\cmslogo}{\includegraphics[height=2.33cm]{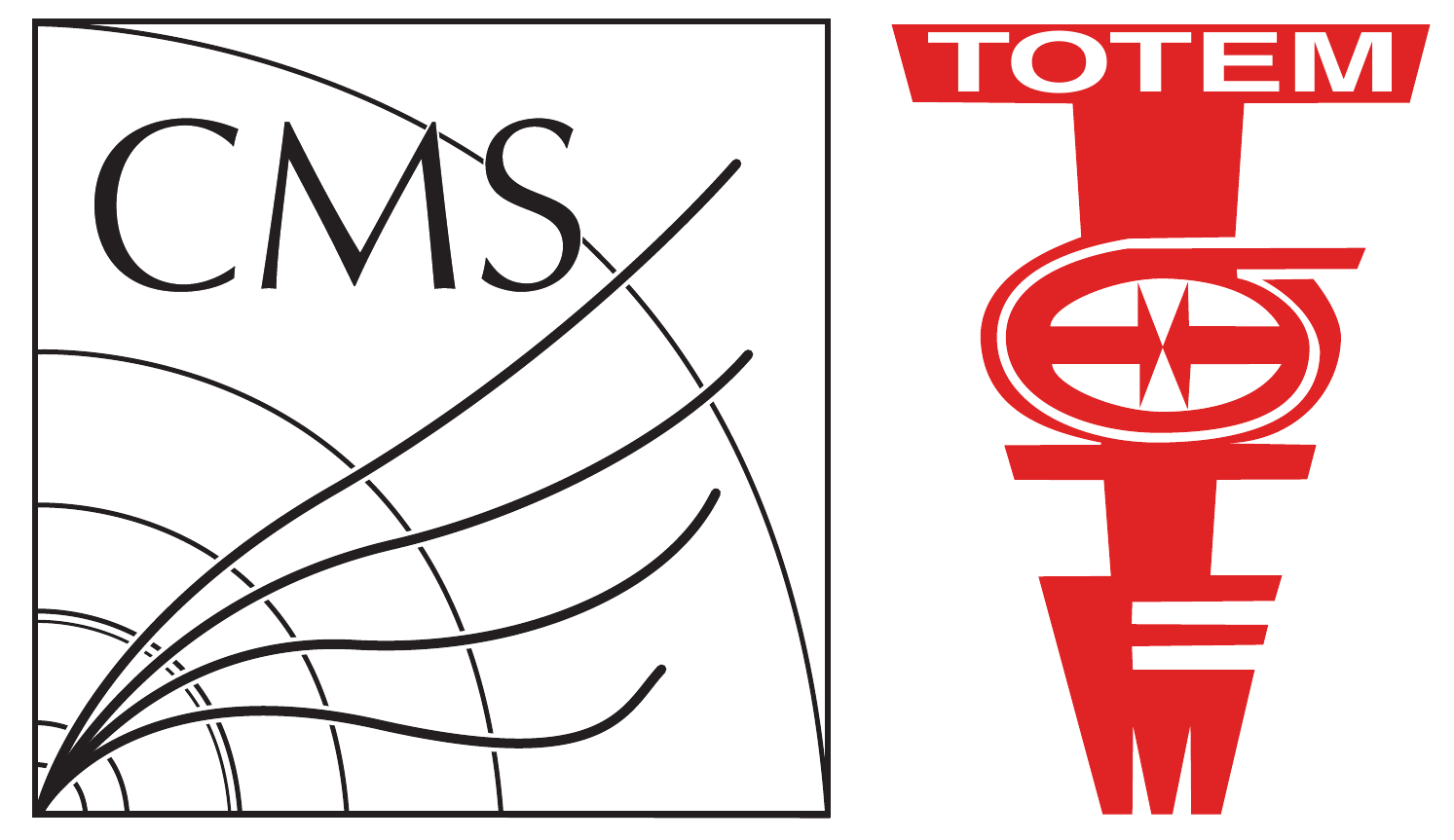}}
\renewcommand{\cmsTag}{CMS-\cmsNUMBER\\&TOTEM-2023-001\\}
\renewcommand{\cmsPubBlock}{\begin{tabular}[t]{@{}r@{}l}&CMS \cmsSTYLE\
\cmsNUMBER\\&TOTEM NOTE 2023-001\\\end{tabular}}
\renewcommand{\cmsCopyright}{\copyright\,\the\year\ CERN for the benefit of the CMS and TOTEM Collaborations.}
\renewcommand{\appMsg}{See Appendix~\ref{app:collab} and \ref{app:totem} for the lists of collaboration members}

\title{A search for new physics in central exclusive production using the missing mass technique with the CMS detector and the CMS-TOTEM precision proton spectrometer}
\titlerunning{A search for new physics with CMS/CMS-TOTEM}
\date{\today}

\abstract{A generic search is presented for the associated production of a \PZ boson or a photon with an additional unspecified massive particle \PX, $\Pp\Pp\to \Pp\Pp +\PZ/\PGg+\PX$, in proton-tagged events from proton-proton collisions at $\sqrt{s}=13\TeV$, recorded in 2017 with the CMS detector and the CMS-TOTEM precision proton spectrometer. The missing mass spectrum is analysed in the 600--1600\GeV range and a fit is performed to search for possible deviations from the background expectation. No significant excess in data with respect to the background predictions has been observed. Model-independent upper limits on the visible production cross section of $\Pp\Pp\to \Pp\Pp +\PZ/\PGg+\PX$ are set.}

\hypersetup{
pdfauthor={CMS Collaboration},
pdftitle={A search for new physics in central exclusive production using the missing mass technique with the CMS-TOTEM precision proton spectrometer},
pdfsubject={CMS and TOTEM},
pdfkeywords={CMS, TOTEM, Z/gamma,  exclusive production, CT-PPS, dark matter}
}

\maketitle

\section{Introduction}
\label{sec:intro}

Despite its remarkable success in describing all known elementary particles and forces, the standard model (SM) of particle physics leaves several fundamental questions unanswered. Such open questions include the nature of dark matter, the asymmetry between matter and antimatter, the hierarchy of particle masses,
and the stability of the Higgs field. Searches for new physics processes beyond the SM (BSM) that might shed light on these issues have so far failed to yield any discoveries. Experiments at the CERN LHC and elsewhere continue to perform new measurements and to cast an even wider net of searches for new phenomena. We do not have clear theoretical indications about the nature of these new phenomena, but they can manifest themselves beyond current theoretical models. 
This motivates a growing interest for model-nonspecific searches~\cite{ATLAS:2018zdn,CMS:2020zjg,ATLAS:2021wob,Karagiorgi:2021ngt} to complement direct model-specific searches.
More comprehensive model-independent frameworks such as effective field theories that probe BSM physics in generic terms~\cite{Buchmuller:1985jz,Grzadkowski:2010es,Ellis:2020unq,Ethier:2021bye,Chapon:2009hh} provide a complementary approach. 
The addition of new detectors, further extending the capability and coverage of the LHC experiments, also offers a new venue to explore final-state topologies and areas of phase space which were not previously covered.

\begin{figure*}[!htp]
\centering
\includegraphics[width=.49\textwidth]{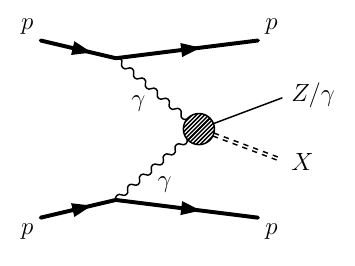}
\caption{
Schematic diagram of the photon-photon production of a \PZ boson or a photon with an additional, unknown particle \PX, giving rise to $\mmiss = \mX$. The production mechanism does not have to proceed through photon exchange. Other colourless exchange mechanisms (\eg double pomeron~\cite{Collins:1977jy}) are also allowed. 
For high-mass central exclusive production, electroweak processes are expected to dominate, and QCD-based colourless exchanges are expected to be suppressed.
}
\label{fig:feyn_diag}
\end{figure*}

This paper describes a generic search for a hypothetical massive particle \PX produced in association with one SM particle in central exclusive production processes at the LHC. In the interaction, the two colliding protons survive after exchanging two colourless particles, and can be recorded in the CMS-TOTEM precision proton spectrometer (CT-PPS hereafter). The detection and accurate measurement of both forward protons allow the full kinematic reconstruction of the event, including the determination of the four-momentum of \PX extracted from the balance between the four-momenta of the tagged SM particle(s) and the forward protons. This ``missing mass" (\mmiss) approach allows searching for BSM particles without assumptions about their decay properties, except that the decay width can be considered narrow enough to produce a resonant mass peak, thus allowing generic BSM searches. 
In this analysis we demonstrate the feasibility of exploiting such a technique 
by searching for a massive particle produced in association with a \PZ boson or a photon in the final state, as illustrated in Fig.~\ref{fig:feyn_diag}.
The analysis relies on the assumptions that the process is exclusive, \ie that nothing is produced in addition to $\PZ/\PGg+\PX$, and that there is no ambiguity in the selection of the charged leptons from the \PZ decay or the \PGg with respect to a possible signature of \PX in the detector.

The \mmiss technique is used to probe the mass of this particle. 
The precise (percent level) proton momentum reconstruction of CT-PPS allows one to search for \mmiss signatures in the high invariant-mass range (600--1600\GeV) covered by the CT-PPS acceptance, with unprecedented resolution. In this high mass range, electroweak processes are generally enhanced relative to quantum chromodynamics (QCD)-induced processes~\cite{Khoze:2001xm} and thus we assume a photon-photon induced exclusive production process.

In 2017, a large sample of proton-proton ($\Pp\Pp$) collision data at $\sqrt{s}=13\TeV$ was recorded with the CMS detector~\cite{Chatrchyan:2008zkk} and CT-PPS~\cite{Albrow:2014lrm}.
The trigger for the central CMS detector is provided by either an isolated photon or a pair of electrons or muons from $\PZ\to\Pell^+\Pell^-$ decays.
The quantity \mmiss is constructed using energy conservation and the four-momenta of the reconstructed boson in the central detector and the final-state protons in CT-PPS.
Final states with a \PZ boson (\PGg) are selected from $\Pp\Pp$ collision data corresponding to an integrated luminosity of 37.2\,(2.3)\fbinv.
Events with a \PGg in the final state were collected with a prescaled trigger.

The paper is organised as follows: the experimental setup is described in Section~\ref{sec:setup}, and it is followed in Section~\ref{sec:physmodel} by the description of the simplified generic Monte Carlo (MC) simulation which is used to simulate the associated exclusive production of a massive particle of narrow width, together with a \PZ boson or photon and forward protons.
The event selection is summarised in Section~\ref{sec:evsel} and the strategy used to estimate the background, which is dominated by a random coincidence of \PZ boson or \PGg production with protons from a different collision in the same bunch crossing (pileup, PU), is discussed in Section~\ref{sec:background}.
The description of the statistical analysis used to compare the observed \mmiss spectrum  with the background plus signal model is presented in Section~\ref{sec:systuncs} and the results are discussed in Section~\ref{sec:results}, followed by a summary in Section~\ref{sec:conclusions}.
Tabulated results are provided in the HEPData record for this analysis~\cite{hepdata}.

\section{The CMS and CT-PPS detectors}
\label{sec:setup}

The central feature of the CMS apparatus is a superconducting solenoid of 6\unit{m} internal diameter, providing a magnetic field of 3.8\unit{T}.
Within the solenoid volume are located a silicon pixel and strip tracker with coverage in pseudorapidity up to $\abs{\eta}=2.5$, surrounded by a lead tungstate crystal electromagnetic calorimeter (ECAL), and a brass and scintillator hadron calorimeter directly outside the ECAL.
The muon detection system consists of three types of gas-ionisation chambers embedded in the steel flux-return yoke outside the solenoid.
A more detailed description of the CMS detector, together with a definition of the coordinate system used and the relevant kinematic variables, can be found in Ref. \cite{Chatrchyan:2008zkk}.
Events are selected online and stored at a maximal rate of about 1 kHz using a two-tier triggering system \cite{Sirunyan:2020zal,Khachatryan:2016bia}.

The CT-PPS is an array of movable, near-beam ``Roman pot" (RP)~\cite{TOTEM:2008lue} devices containing tracking and timing detectors inserted horizontally at a distance from the beam corresponding to about $14$ standard deviations of the transverse distribution of the LHC beam, \ie in the range from 1.5 to 3.0\mm.
The detectors are used to reconstruct the flight path of protons coming from the interaction point (IP) through 210 m of LHC beamline.

The tracking stations provide a measurement of the proton trajectories with respect to the beam position.
Knowledge of the magnetic fields traversed by the proton from the IP to the RPs allows the reconstruction of its fractional momentum loss, with respect to the momentum of the incident proton, $\xi=\Delta p/p\sim D_x^{-1} x$, where $x$ represents the horizontal displacement of the scattered proton at the RP location, and $D_x$ the horizontal dispersion, a property of the accelerator optics.
The techniques used for the alignment and calibration of the apparatus and proton reconstruction are detailed in Refs.~\cite{TOTEM:2022vox, CMS-DP-2020-047,CMS-DP-2019-035,CMS-DP-2019-036}.

In its 2017 configuration, CT-PPS is able to measure protons which have lost approximately 2-20\,\% of their initial momentum,
and which remain inside the beam pipe until they hit the RPs. This corresponds to an invariant mass of the central system in the range 600--1600\GeV. 

The performance of CT-PPS and its potential for high-mass exclusive measurements was validated by the observation, by the CMS and TOTEM Collaborations, of proton-tagged (semi-) exclusive dilepton events recorded in 2016~\cite{Cms:2018het}.

If both the forward protons and the central boson are measured, the kinematics of the hypothetical \PX particle can be fully reconstructed. Given the excellent resolution of the charged-lepton and photon reconstruction in the central CMS detector and of the protons in CT-PPS, we search for a resonance in the \mmiss distribution.

The missing mass is defined as:
\begin{equation}
\mmiss^2=\left[(P^\text{in}_{\Pp_1}+P^\text{in}_{\Pp_2}) - (P_\PV+P^\text{out}_{\Pp_1}+P^\text{out}_{\Pp_2})\right]^2,
\label{eq:mmass}
\end{equation}
where $P_\PV$ is the four-momentum of the boson and $P^\text{out,in}_{\Pp_i}$ ($i=1$, 2) are the four-momenta of the outgoing and incoming protons, respectively.

In this analysis, two RP tracking stations per side, or ``arm'', of CMS are used. The stations in LHC sector 56 are located on the side that corresponds to positive $z$ coordinates in the CMS coordinate system, while the stations on the other side are located in LHC sector 45, corresponding to negative $z$.      
In 2017, the inner stations, closest to the CMS detector, were instrumented with silicon strip detectors and the outer stations with 3D pixel detectors.
The strip detectors were designed for low-luminosity and low-PU conditions. They cannot resolve multiple tracks and have lower radiation hardness. In the case of multiple protons crossing a strip detector, zero tracks are reconstructed. In spite of these limitations, the reconstruction of protons requiring tracks in both the pixel and strip detectors (multi-RP reconstruction) yields high-purity samples of protons with excellent momentum resolution, which more than compensates for the loss in the overall acceptance and efficiency.
After comparing various reconstruction options, the multi-RP reconstruction was selected as the preferred method.
However, to maximise the efficiency of this search, pixel-only reconstruction (single-RP) is used as a fallback in case the multi-RP reconstruction fails in an event.

\section{Physics model and event simulation}
\label{sec:physmodel}

\subsection{Signal model}
\label{subsec:signalmodel}

The signal is simulated using a simplified MC model that uses as main inputs the mass of the \PX particle, $\mX$, and the $p_z$ spectrum of the $\PV\PX$ system, where $\PV=\PZ,\PGg$.
It is assumed that the vector boson \PV is produced isotropically in the $\PV\PX$ frame, and that the leptons from the $\PZ$ decay are also produced isotropically in the $\PZ$ reference frame. Samples are produced for a range of $\mX$ values, with the requirement that the corresponding mass of the $\PV\PX$ system ($m_{\PV\PX}$) falls within the acceptance of CT-PPS. For each event, the $m_{\PV\PX}$ is generated assuming an exponential spectrum, $m_{\PV\PX}=\mX+\varepsilon+100\GeV$, where $\varepsilon$ is a randomly distributed variable following an exponential probability distribution function with decay constant $\tau=-0.04\GeV^{-1}$.
The values chosen for these constants were found to provide a good coverage of the phase space within the CT-PPS acceptance; however, their precise values and the choice of the shape of the spectrum have negligible effect on the result of this analysis.
The di-proton $p_z$ values, $p_z(\Pp\Pp)$, on the other hand, have a sizeable effect on the proton acceptance. They are, by momentum conservation, the same as those of the $\PV\PX$ system but have opposite sign; the $p_z$ distribution is obtained using the equivalent photon approximation~\cite{Budnev:1975poe} resulting in a Gaussian distribution with a mass-dependent width, ranging from 856 to 1250 GeV for \PX masses varying from 600 to 1600 GeV.

The outgoing protons are transported from the IP to the CT-PPS RPs using the LHC optics, taking into account aperture limitations, which depend on the beam crossing angle. The simulation of the protons includes beam divergence and vertex smearing at the IP also depending on the beam crossing angle, as detailed in Ref.~\cite{CMS-DP-2020-047}.
The hits in the RP detectors are generated taking into account sensor acceptance, efficiency, and resolution.
Corrections for the acceptance and reconstruction efficiency of the photon or the leptons from Z boson decay in the central CMS detector are also applied. More details are given in Section~\ref{sec:evsel}.

Given the wide $p_z$ spectrum of the $\PV\PX$ system with respect to the CT-PPS acceptance,
we divide the generated events into two sets: inside and outside the fiducial volume.
The definition of the fiducial volume is based on the kinematic properties of the generator-level particles from the hard process:
the boson (or its decay products) and the outgoing protons are used in the definition as summarised in Table~\ref{tab:fidcuts}.
The requirements on the outgoing protons are not symmetric.
The acceptances of the $+z$ and $-z$ arms of CT-PPS are different, mostly as a result of differences in the LHC optics settings. The requirements for the $\xi$ range at generator level are chosen accordingly for the proton in the +z arm ($\xi_{+}^\text{gen}$) and -z arm ($\xi_{-}^\text{gen}$). Four main LHC beam crossing angle configurations are used in 2017, each resulting in a different detector acceptance as a function of $\xi$.
The average of the lower edges of the $\xi$ acceptance is defined as the start of the $\xi$ range for the fiducial region, and the average of the upper edges as the end of the range~\cite{TOTEM:2022vox}. This approach offers a compromise between maximal acceptance of the protons recorded in the data and minimal use of outer regions of phase space that were not fully efficient during the entire data taking due to varying beam configurations.
The fiducial volume is used to define the reference normalisation of the signal.

Events falling outside of these fiducial requirements are considered as background in this analysis; the normalisation of the background is allowed to float freely in the fits described later, independently of the signal term. 
This definition reduces the dependence on model assumptions, and specifically the precise shape of the $p_z$ spectrum of the generated signal.

\begin{table*}[!htp]
\centering
\topcaption{ Combined CMS+CT-PPS fiducial volume selection criteria in the \PZ and \PGg analyses.
The leading- and subleading-\pt leptons,
where \pt is transverse momentum,
are labelled as $\Pell_1$ and $\Pell_2$, respectively.
The \PZ boson mass is noted as $m_{\PZ}$.
\label{tab:fidcuts} }
\begin{tabular}{ lcc }
\hline
Selection/analysis & $\PZ\to \Pep\Pem$/$\PZ\to\Pgm^+\Pgm^-$ & \PGg \\[4pt] \hline
   & ~~ ${\geq}2$ same-flavour leptons (\Pe{} or \Pgm) ~~ & \\[4pt]
                & opposite electric charge                      & \\[4pt]
Leptons/photons ~~~~ & $\pt(\Pell_1)>30\GeV,~\abs{\eta(\Pell_1)}<2.4$   & ~~~~ 1\PGg within $\abs{\eta(\PGg)}<1.44$ \\[4pt]
                & $\pt(\Pell_2)>20\GeV,~\abs{\eta(\Pell_2)}<2.4$   & \\[4pt]
                & $\abs{m(\Pell_1,\Pell_2)-m_{\PZ}}<10\GeV$         & \\[4pt]
Boson \pt       & $\pt(\PZ)>40\GeV$                             & $\pt(\PGg)>95\GeV$ \\[4pt]
Protons & \multicolumn{2}{c}{~~~~~~~~~~~~~~$0.02<\xi_{+}^\text{gen}<0.16$ ~ and ~ $0.03<\xi_{-}^\text{gen}<0.18$} \\[4pt]
\hline
\end{tabular}
\end{table*}

\subsection{Simulation samples for background validation}
\label{subsec:background}

Although the background in this analysis is fully modelled from data,
standard full simulation MC samples, including the simulation of the CMS detector response with \GEANTfour~\cite{AGOSTINELLI2003250},
are used to optimise the event selection and to validate the efficiencies of the particle reconstruction and identification, as well as of the triggers. Each background process is generated in coincidence with additional minimum-bias events using {\PYTHIA}8 to simulate PU events, and with a frequency distribution matching that observed in the data. 
The PU protons are modelled with a dedicated event mixing technique, which is explained in more detail in Section~\ref{sec:background}.

The main process considered for the validation is Drell-Yan, in particular \PZ boson production ($\PZ+$jets), simulated using
\MGvATNLO v2.2.2, with FxFx merging~\cite{Frederix:2012ps}, at next-to-leading order (NLO) precision in QCD.
For the photon analysis the main process considered 
 is the production of an isolated photon in association with jets
($\PGg+$jets), modelled at the leading-order using \MGvATNLO with MLM merging~\cite{Alwall:2007fs}.
Top quark production processes (single top $\PQt\PW$ and \ttbar) are also simulated with
\POWHEG~\cite{Re:2010bp,Frixione:2007nw} as well as diboson production ($\PW\PW$, $\PW\PZ$, and $\PZ\PZ$),
the latter using {\PYTHIA}8 version 8.226~\cite{Sjostrand:2014zea}.
The {\PYTHIA}8 program is used as the parton shower generator for the processes simulated with a matrix element approach.
Furthermore, single- and double-diffractive~\cite{Arneodo:2005kd} \PZ boson samples are produced with {\PYTHIA}8 and \textsc{Pomwig}~\cite{Cox:2000jt} to check that the contribution of these processes to the final sample is consistent with the background estimate based on the data.

\section{Event reconstruction and selection}
\label{sec:evsel}

A particle-flow algorithm~\cite{CMS-PRF-14-001} is used for offline reconstruction of the events in the central CMS detector, exploiting an optimised combination of information from the various subdetectors.
The energy of photons is obtained from the ECAL measurement,
while the energy of electrons is determined from a combination of the electron momentum at
the primary interaction vertex as determined by the tracker,
the energy of the corresponding ECAL cluster,
and the energy sum of all bremsstrahlung photons spatially compatible with originating from the electron track~\cite{CMS:2020uim}.
The energy of muons is obtained from the curvature of the corresponding tracks~\cite{CMS:2019ied}.

Protons coming from the IP are reconstructed by CT-PPS, either from a single RP station (single-RP reconstruction) or using two RP stations (multi-RP reconstruction). In the single-RP method, only pixel detectors are used for proton reconstruction since the strip detectors have significantly lower efficiency. Efficiency corrections are applied to the simulated protons and account for three components: the effect of radiation damage in the RP strip and pixel detectors, the inefficiency of the strip stations when they are hit by several protons, and the effect of matching the pixel and strip stations in a combined multi-RP proton fit. The radiation damage efficiency corrections are parametrised as functions of time and beam crossing angle, separately for the strip and pixel detectors in each arm of CT-PPS. All these efficiencies have been measured in data. The efficiency of matching the pixel and strip stations in the combined multi-RP proton reconstruction has also been measured in data. Further details can be found in Ref.~\cite{TOTEM:2022vox}.

The events are selected if they pass a single-photon trigger
(transverse momentum $\pt>90\GeV$ in the ECAL barrel region), or a combination of double-lepton triggers with the leading lepton $\pt >17\,(23)\GeV$ and the subleading lepton $\pt >8\,(12)\GeV$ for muons (electrons) and a single-muon trigger ($\pt>27\GeV$).
Because of its larger rate, the photon trigger was prescaled throughout the data taking period analysed,
resulting in a data set with lower effective integrated luminosity.

Offline, the events are preselected if they contain one isolated photon with $\pt>95\GeV$ in the barrel,
or at least two leptons passing the electron or muon identification and isolation criteria.
In all cases the isolation is based on the \pt of all particle-flow candidates found in a cone around the object being selected.
An isolation cone with a radius of $\sqrt{\Delta\eta^2+\Delta\phi^2}=0.3$, where $\phi$ is the azimuthal angle in radians, is used for electrons and photons, and 0.4 for muons.

For the leptonic selection, the following requirements must be fulfilled:
\begin{itemize}
\item leading lepton: $\pt>30\GeV$ and $\abs{\eta}<2.4$;
\item sub-leading lepton: $\pt>20\GeV$ and $\abs{\eta}<2.4$;
\item dilepton mass $m(\Pell_1, \Pell_2)>20\GeV$, and the two leptons have opposite electric charge.
\end{itemize}

The events are then categorised as:
\begin{itemize}
\item Same-flavour events ($\Pe\Pe$ or $\Pgm\Pgm$) --
      if $m(\Pell_1, \Pell_2)$ is consistent with the mass of a decayed \PZ boson, \ie $\abs{m(\Pell_1,\Pell_2)-m_\PZ}<10\GeV$.
      These events are furthermore categorised in a control region with $\pt(\PZ)<10\GeV$ and a search region with $\pt(\PZ)>40\GeV$;
\item Single-photon events (\PGg) -- if one photon is selected at the trigger and offline levels;
\item Different-flavour events ($\Pe\Pgm$) -- used as a control sample for background modelling.
\end{itemize}

No veto requirements for additional objects in the event are imposed, thus allowing the presence of other activity in the event besides the leptons from the \PZ boson decay, or the photon.
In the signal regions of the same-flavour and single-photon samples, in addition to the selection criteria described above,
it is required that at least one proton be reconstructed in each arm of CT-PPS.
The protons are obtained from the combined multi-RP reconstruction (default), or the pixel-only single-RP reconstruction
when the former is not available in an event. Dedicated calibrations are used for the two main periods of the 2017 data taking,
and for the 120, 130, 140, and 150 $\mu$rad LHC beam crossing angles. Events recorded during periods with other angles are excluded from this analysis.
The reconstructed protons are pre-selected with $\xi_{\pm}>0.035$.
Additional requirements on the lower value of $\xi$ remove the region where the acceptance varies quickly close to the edge of the sensors; likewise an upper $\xi$ requirement is imposed to reflect the LHC collimator apertures.
Further details can be found in Ref.~\cite{CMS-DP-2020-047}.

The simulation of the signal samples is performed at the generator level and does not include the CMS detector response. The efficiency and acceptance corrections are applied to the generated leptons from the \PZ boson decay, or to the generated central photon, based on the expectations from CMS full simulation MC samples with similar dilepton/photon topologies. As a cross-check, for one of the mass points ($\mX=950\GeV$), the full CMS detector simulation and PU conditions are applied and used for comparison. The results of the full and the simplified simulation are in good agreement.

\begin{figure*}
\centering
\includegraphics[width=.35\textwidth]{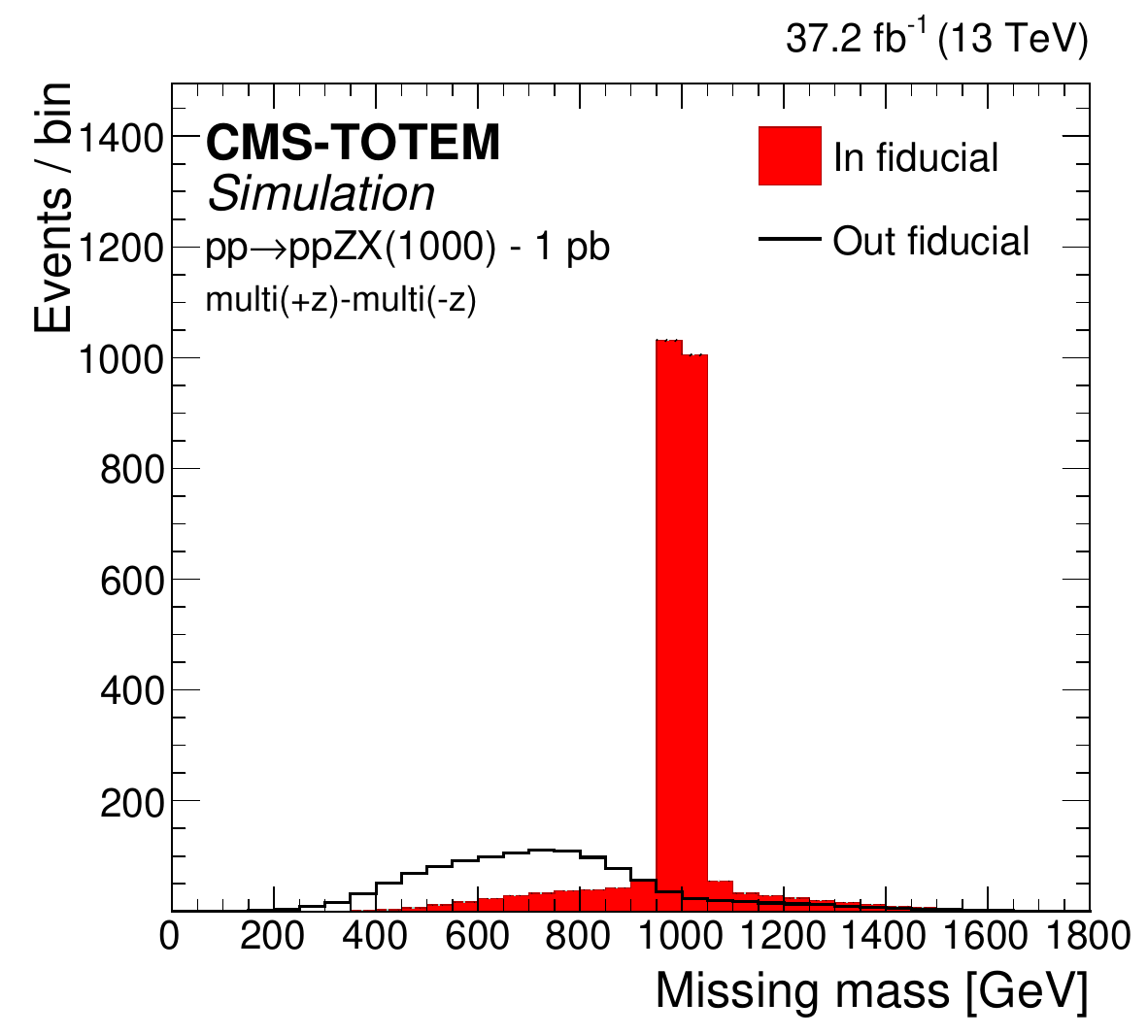}
\includegraphics[width=.35\textwidth]{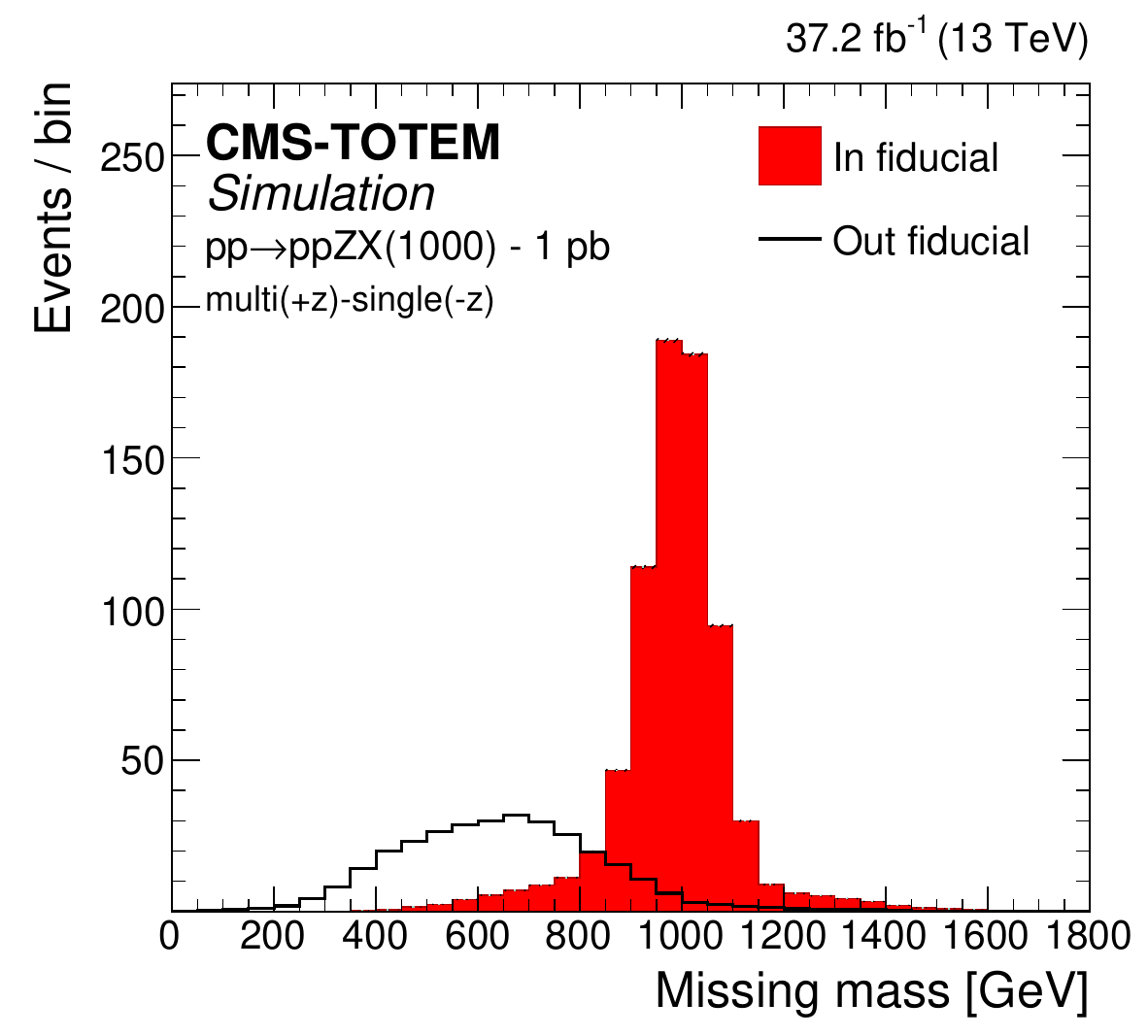}\\
\includegraphics[width=.35\textwidth]{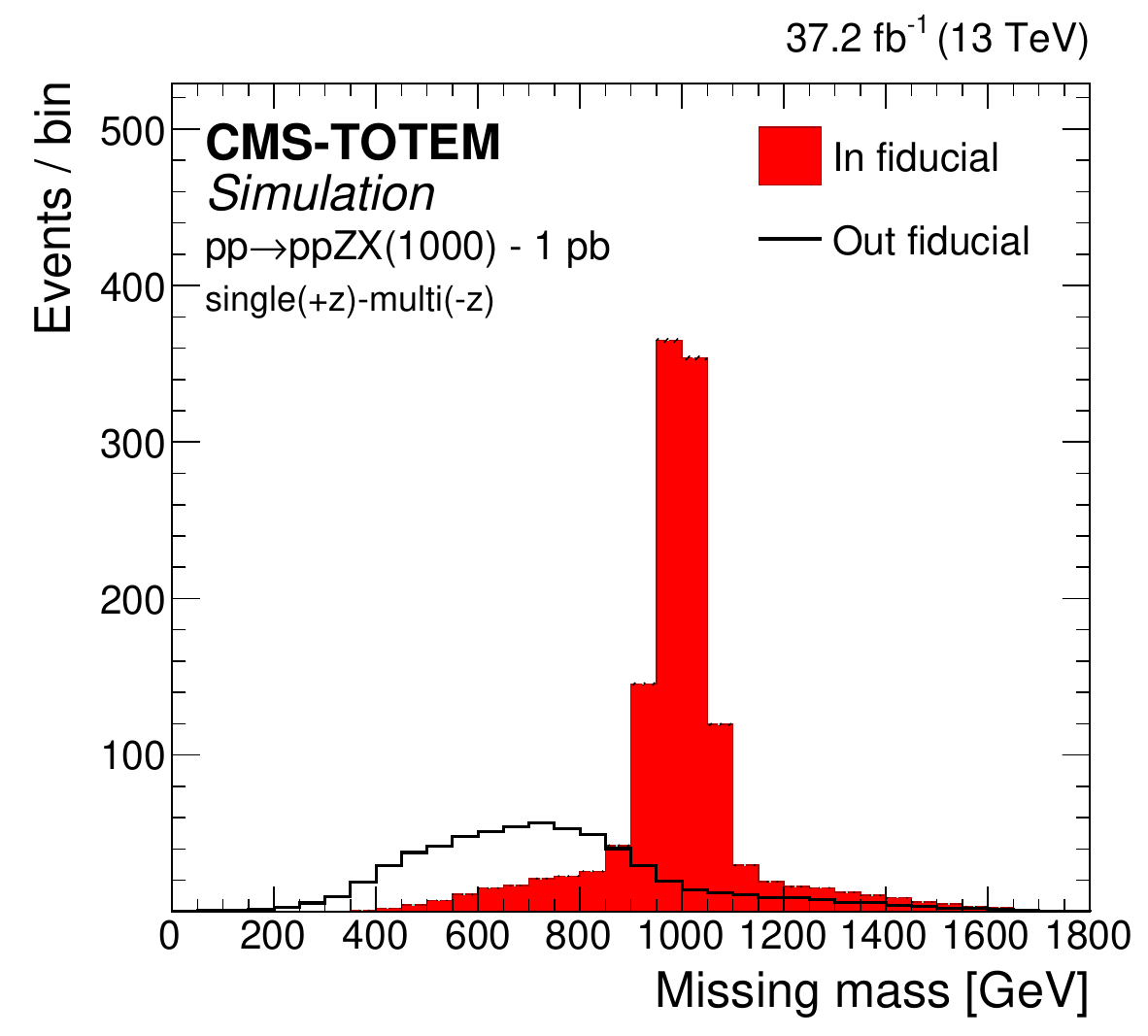}
\includegraphics[width=.35\textwidth]{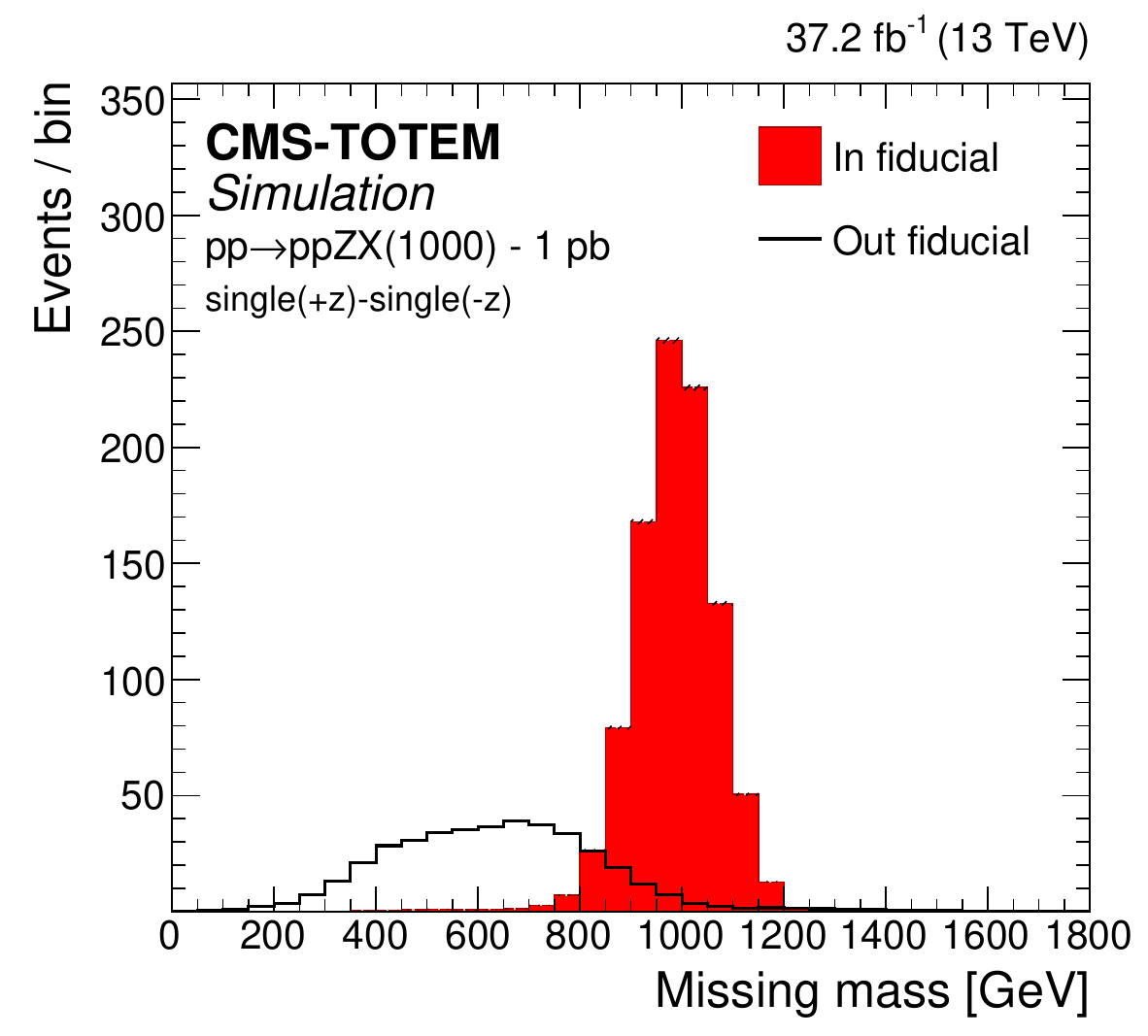}
\caption{
Comparison of the \mmiss shapes for the simulated $\Pp\Pp\to \Pp\Pp\PZ\PX$ signal events within the fiducial region and those outside it, after including the effect of PU protons as described in the text, for a generated $\mX$ mass of 1000\GeV. A fiducial cross section of 1\unit{pb} is used to normalise the simulation.
From left to right and top to bottom, the distributions are shown for the different proton reconstruction categories: multi($+z$)-multi($-z$), multi($+z$)-single($-z$), single($+z$)-multi($-z$) and single($+z$)-single($-z$).}
\label{fig:mmass_signal_out}
\end{figure*}

Figure~\ref{fig:mmass_signal_out} shows the expected signal shape as a function of \mmiss for events inside and outside the fiducial region for a generated $\mX=1000\GeV$ and for different categories of proton reconstruction: a multi-RP proton on both sides, one side, or pixel-only single-RP protons on both sides. One can observe that
events outside the fiducial region, where at least one signal proton is not in the CT-PPS acceptance, are expected to have a non-resonant spectrum. In addition there is a clear difference in shape between the two components, and the resolution of the peak for events in the fiducial region depends strongly on the proton reconstruction used.
The expected resolution in the reconstruction of the \mmiss variable is $\approx$2\% (7\%) with the multi- (single-)RP  reconstruction algorithm. The so-called fallback categories, which make use of pixel-only reconstruction, have a worse resolution and also contain a smaller number of selected events with respect to the main category in this analysis. Nevertheless, these categories contribute to increasing the overall sensitivity of the search by $\approx 20\%$.
Events inside the fiducial region exhibit also a non-resonant component because of PU. The shape and relative contribution of this component is category-dependent because of the different reconstruction efficiency and acceptance of the detectors.

\begin{figure*}
\centering
\includegraphics[width=.49\textwidth]{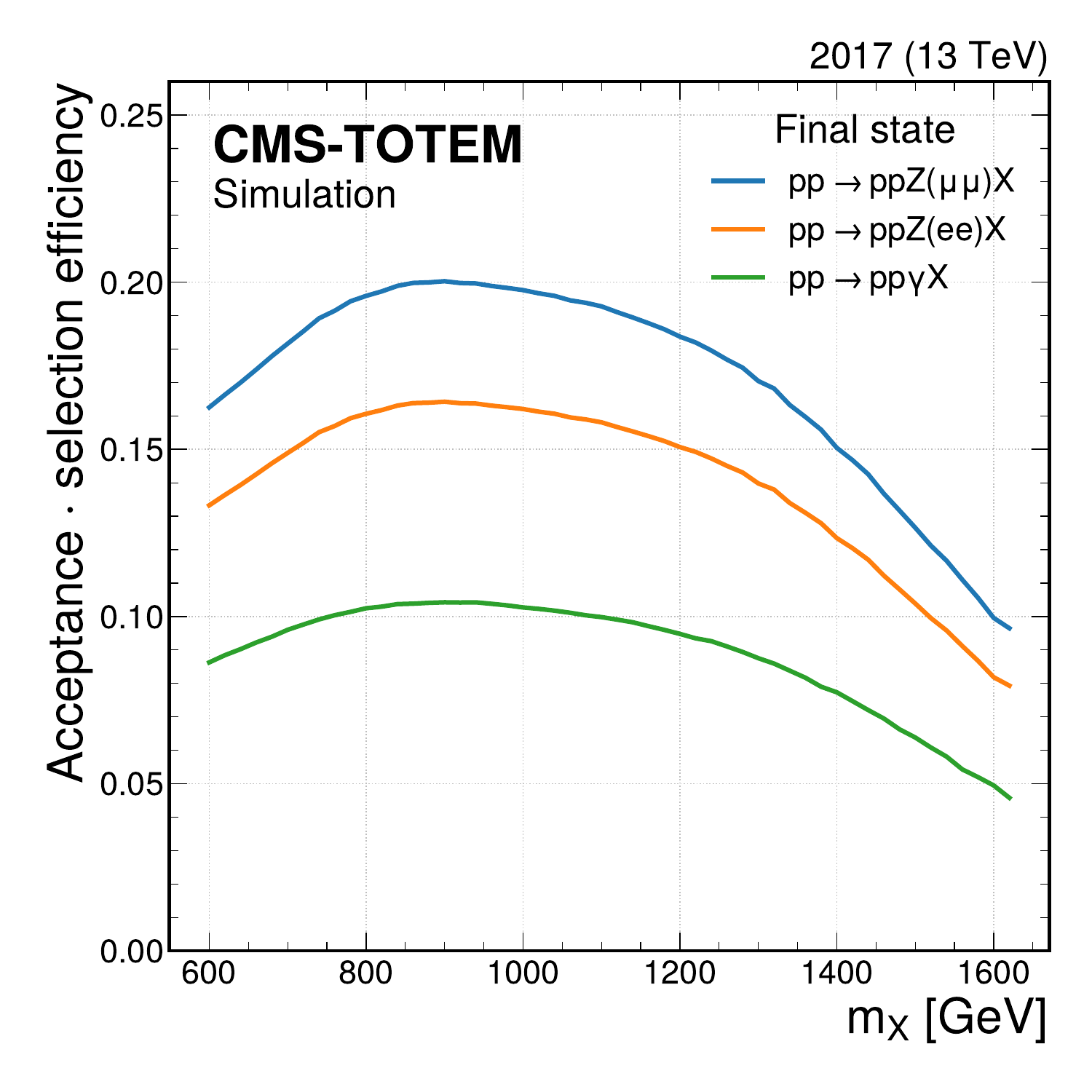}
\includegraphics[width=.49\textwidth]{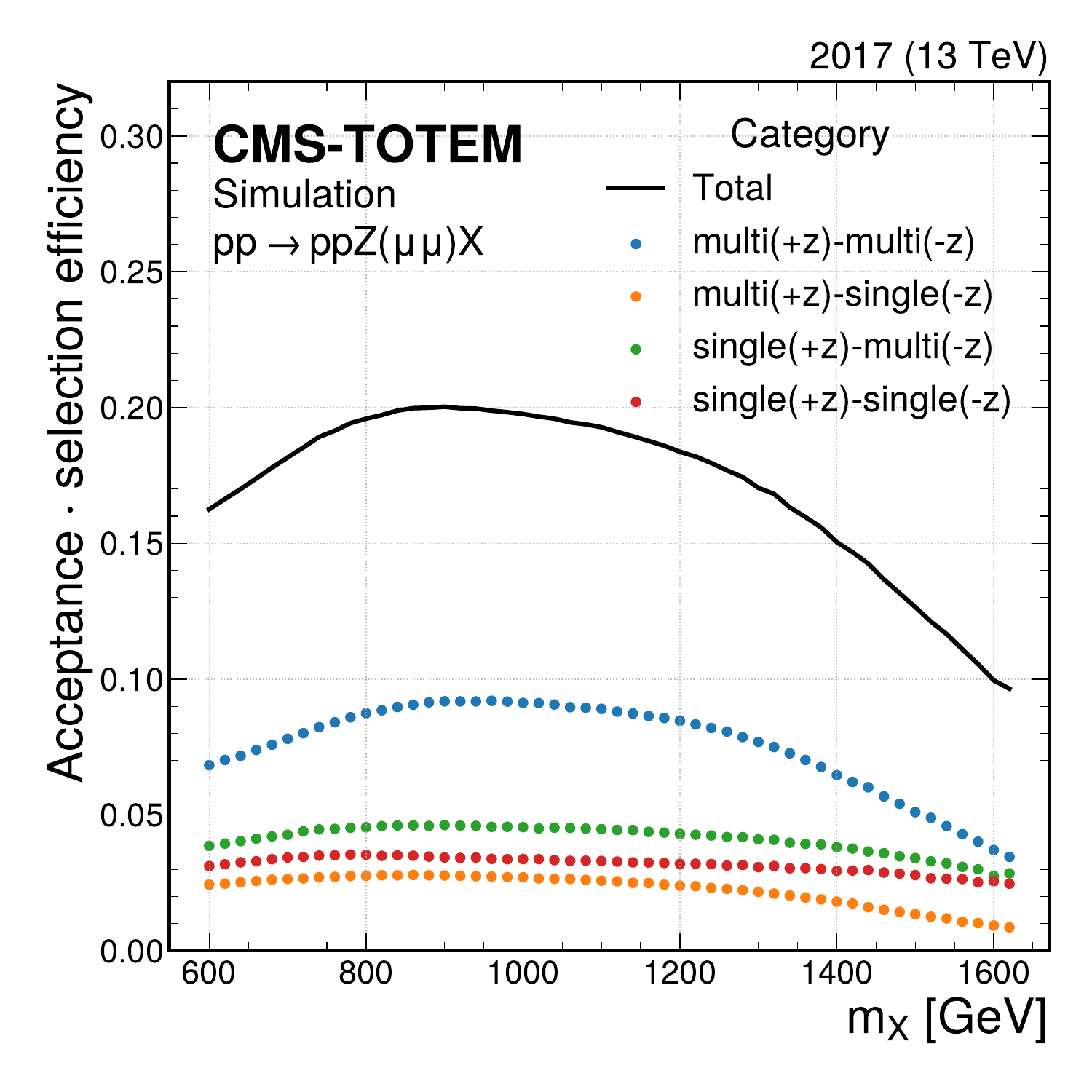}
\caption{Product of the acceptance and the combined reconstruction and identification efficiency, as a function of $\mX$, for events generated inside the fiducial volume defined in Table~\ref{tab:fidcuts}.
The curves shown in the left panel display the different final states, 
while the ones in the right panel show the contributions from the different proton reconstruction categories
in the $\PZ\to\Pgm\Pgm$ analysis: multi($+z$)-multi($-z$), multi($+z$)-single($-z$), single($+z$)-multi($-z$) and single($+z$)-single($-z$). 
The definitions of the fiducial region and of the signal model used to estimate the acceptance are provided in the text.}
\label{fig:analysis_acceptance}
\end{figure*}

Figure~\ref{fig:analysis_acceptance} summarises the overall product of acceptance and selection efficiency for the \PZ and \PGg analyses including both the central object(s) (photon or leptons) and the proton contributions, for events produced inside the fiducial volume.
In both channels, the overall product of acceptance and efficiency is maximal for $\mX$ in the range 800--1150\GeV. For about half of the events, the multi-RP reconstruction algorithm is used for both protons. An important contribution to the loss of the efficiency for multi-RP protons in the 2017 data is due to the inclusion of downtime of parts of the detector as an inefficiency component
and to the  inefficiency caused by multiple tracks from PU protons in the RP strip-detectors~\cite{TOTEM:2022vox}. Approximately a factor of two in overall efficiency is recovered through the inclusion of pixel-only single-RP protons when multi-RP protons are not available.

\section{Background model}
\label{sec:background}

Several sources of backgrounds are considered:

\begin{itemize}
    \item inclusive SM processes (mainly $\PZ+$jets or $\PGg+$jets), with two protons from PU events (combinatorial background);
    \item a single-diffractive process, with one additional proton from a PU event;
    \item double-diffractive processes;
    \item exclusive SM processes (mainly $\PGg\PGg\to\Pell\Pell$);
    \item signal-induced background (from signal events in which one or both of the protons escape detection).
\end{itemize}

Of these sources, by far the most important one is the combinatorial background,
originating from the random superposition of proton candidates from PU collisions
reconstructed in the RPs on both sides.
We employ a background model fully based on data, by replacing the proton on one side of the CMS detector (single mixing)
or on both sides (double mixing), with a proton (or protons) from a randomly chosen event, from the same data taking period and with the same beam conditions. It is expected that the background model correctly reproduces the combinatorial background by mimicking precisely the process of the superposition of uncorrelated pile-up protons, and that the single mixing simulates to good approximation the contribution from single-diffractive processes.
The contributions from other fully exclusive or double-diffractive processes are estimated, from simulation, to be negligible.
The signal-induced background is taken from the signal simulation and included separately in the final fit.

In order to avoid any residual statistical correlation with the hypothetical signal present in the signal region,
the protons for the mixing procedure are taken from a control sample that is orthogonal by construction to the signal sample.
This is done by using $\PZ\to\Pgm\Pgm$ events with $\pt(\PZ)<10\GeV$.
The background estimate obtained with protons from a different control region ($\Pe\Pgm$ events passing the full selection) is compared
to the default background model, and the difference considered as a systematic uncertainty.
An additional systematic uncertainty is considered from the difference found with respect to the background estimate obtained from single mixing,
reflecting a possible difference between the modelling of purely combinatorial background and single-diffractive processes.

The mixing with random protons from other events is repeated 100 times per data event in order to increase the statistical precision of the background prediction.
The same procedure is also applied to simulated events, taking care that the relative fractions of integrated luminosity for each of the data-taking periods is respected when selecting the protons to be mixed.

\begin{figure*}[!htp]
\centering
\includegraphics[width=.32\textwidth]{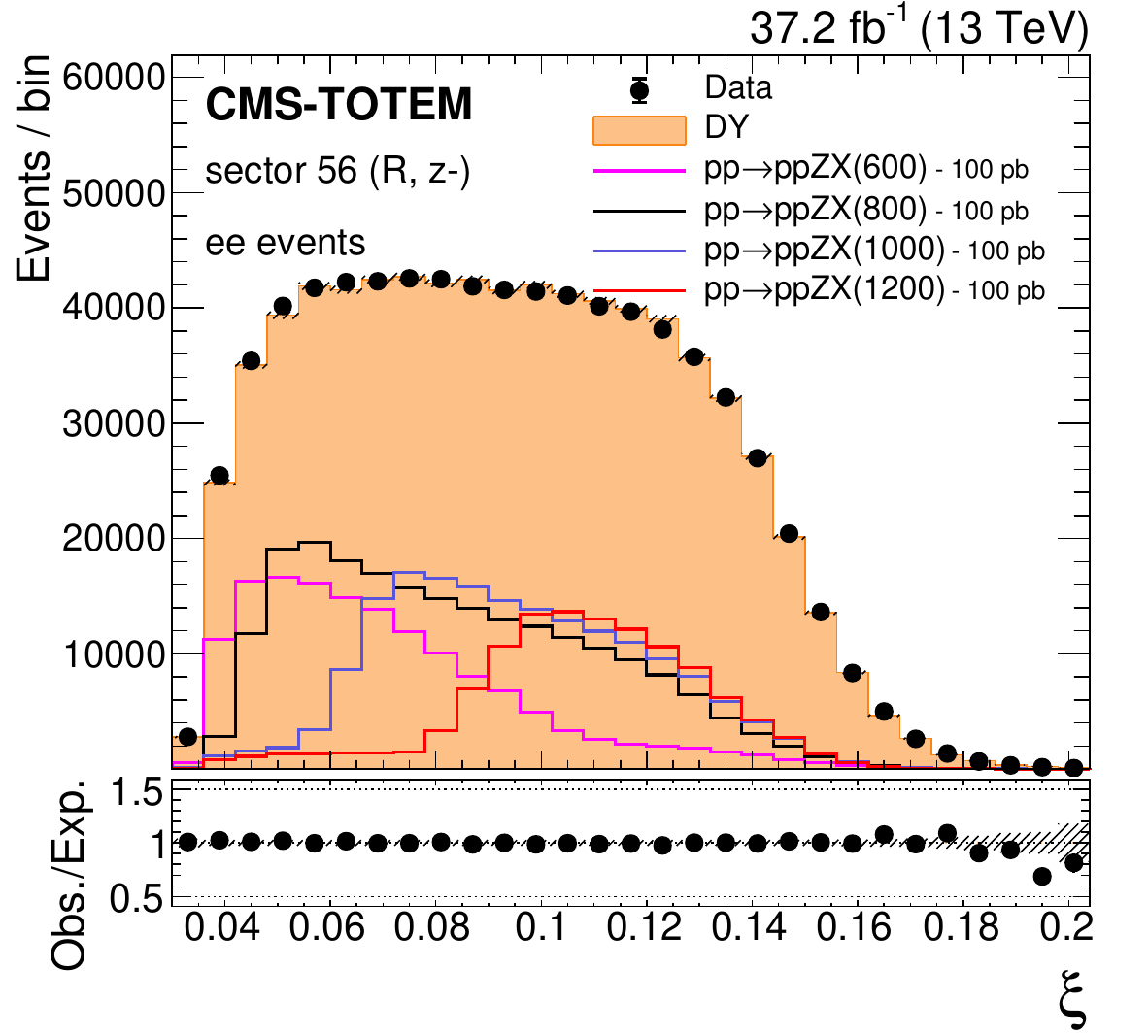}
\includegraphics[width=.32\textwidth]{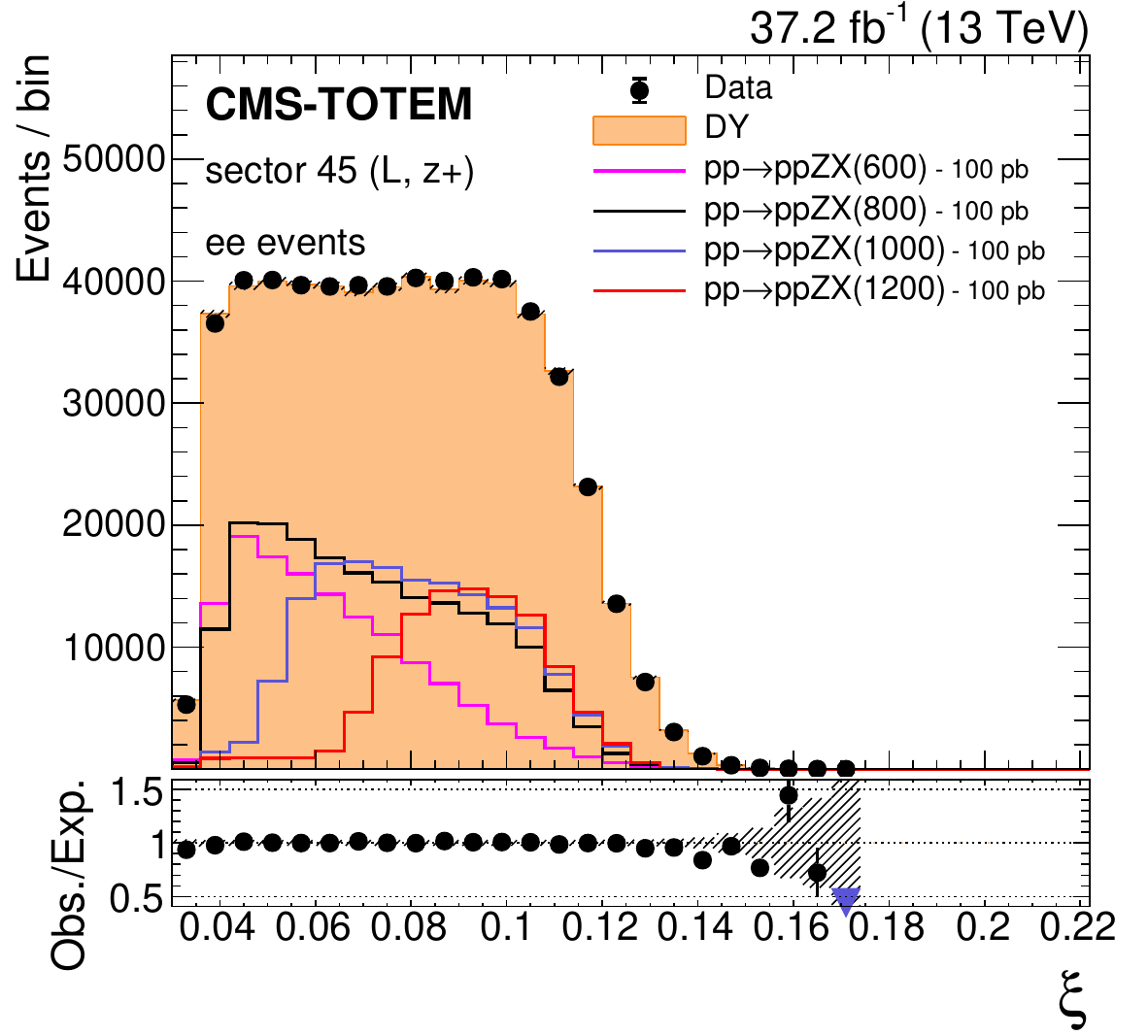}
\includegraphics[width=.32\textwidth]{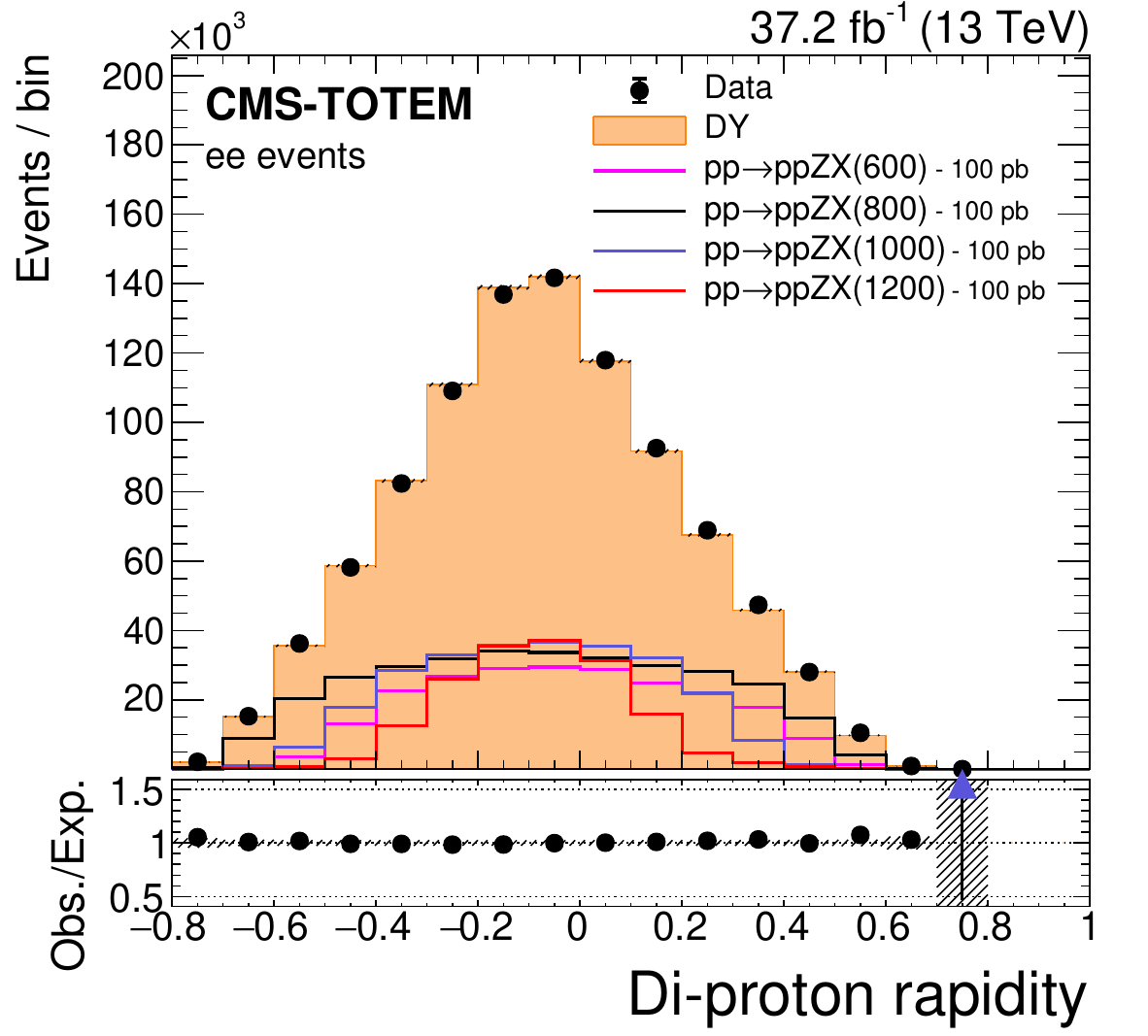}\\
\includegraphics[width=.32\textwidth]{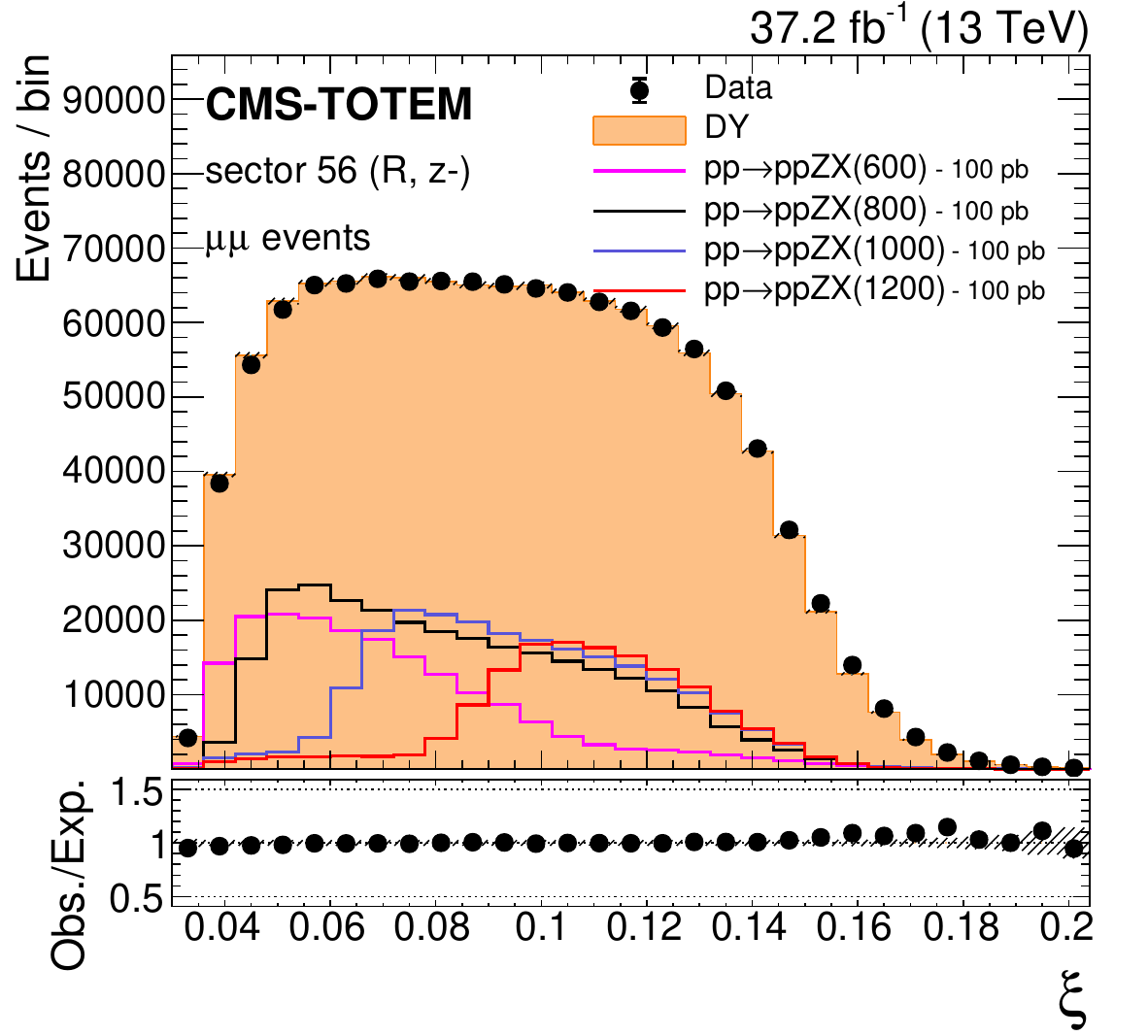}
\includegraphics[width=.32\textwidth]{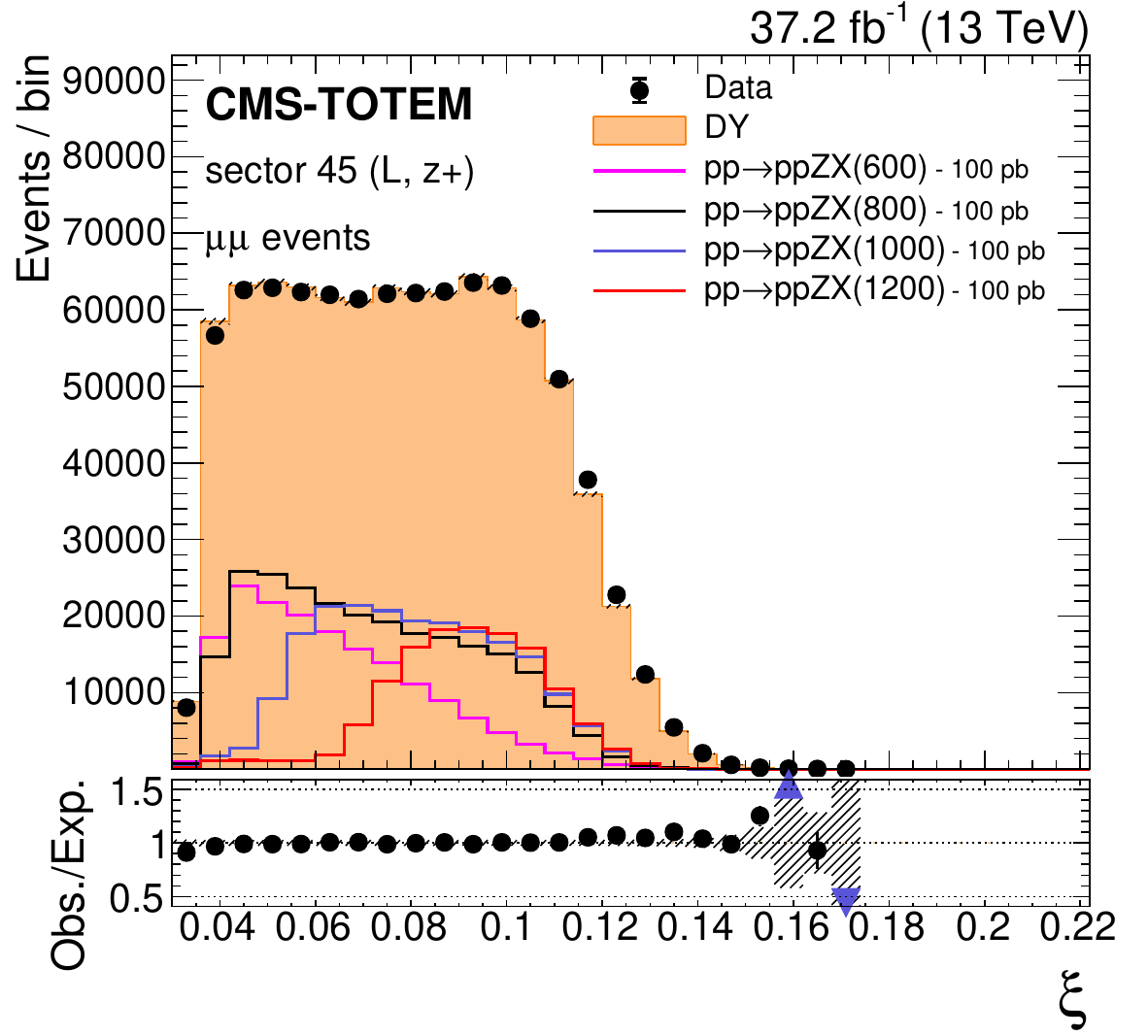}
\includegraphics[width=.32\textwidth]{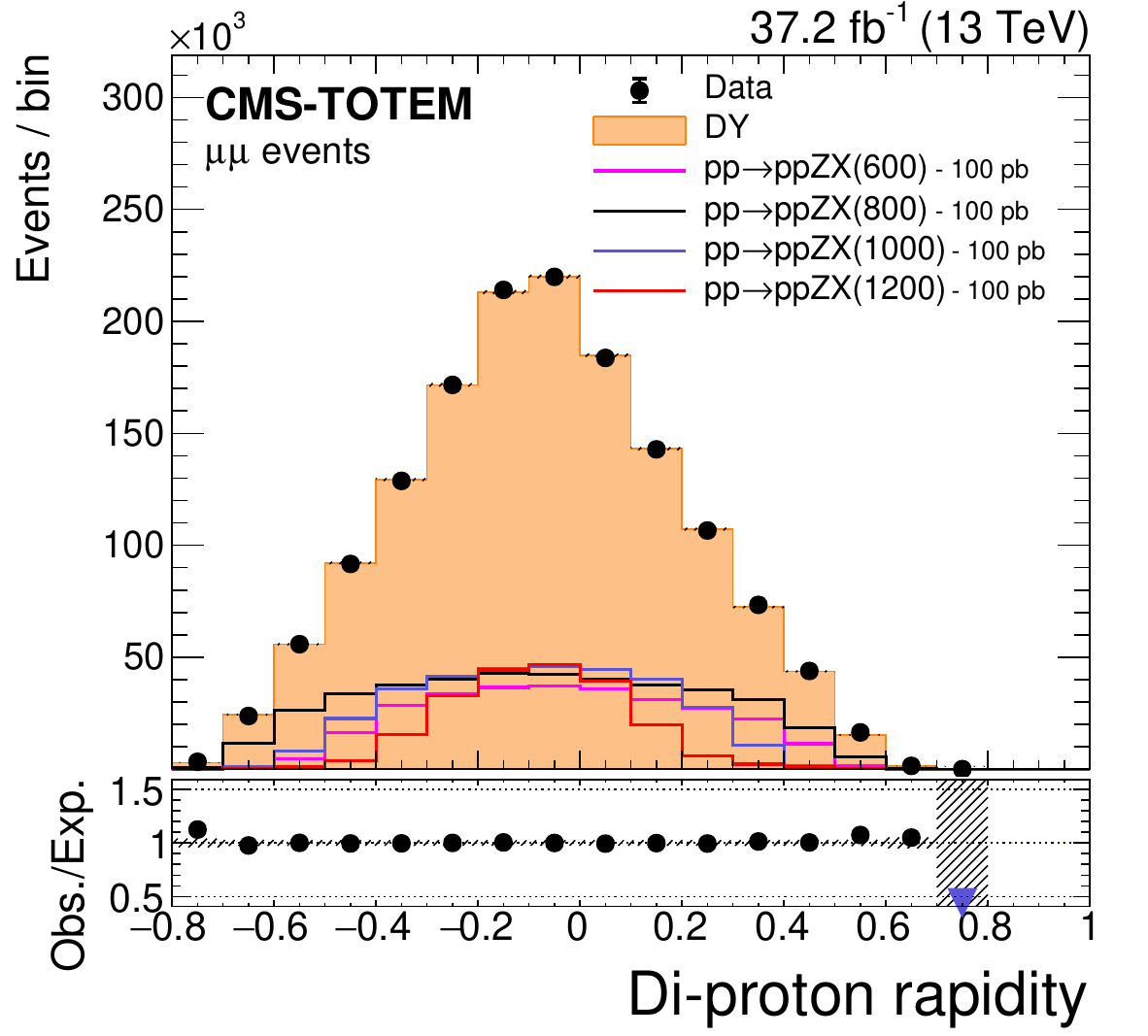}\\
\includegraphics[width=.32\textwidth]{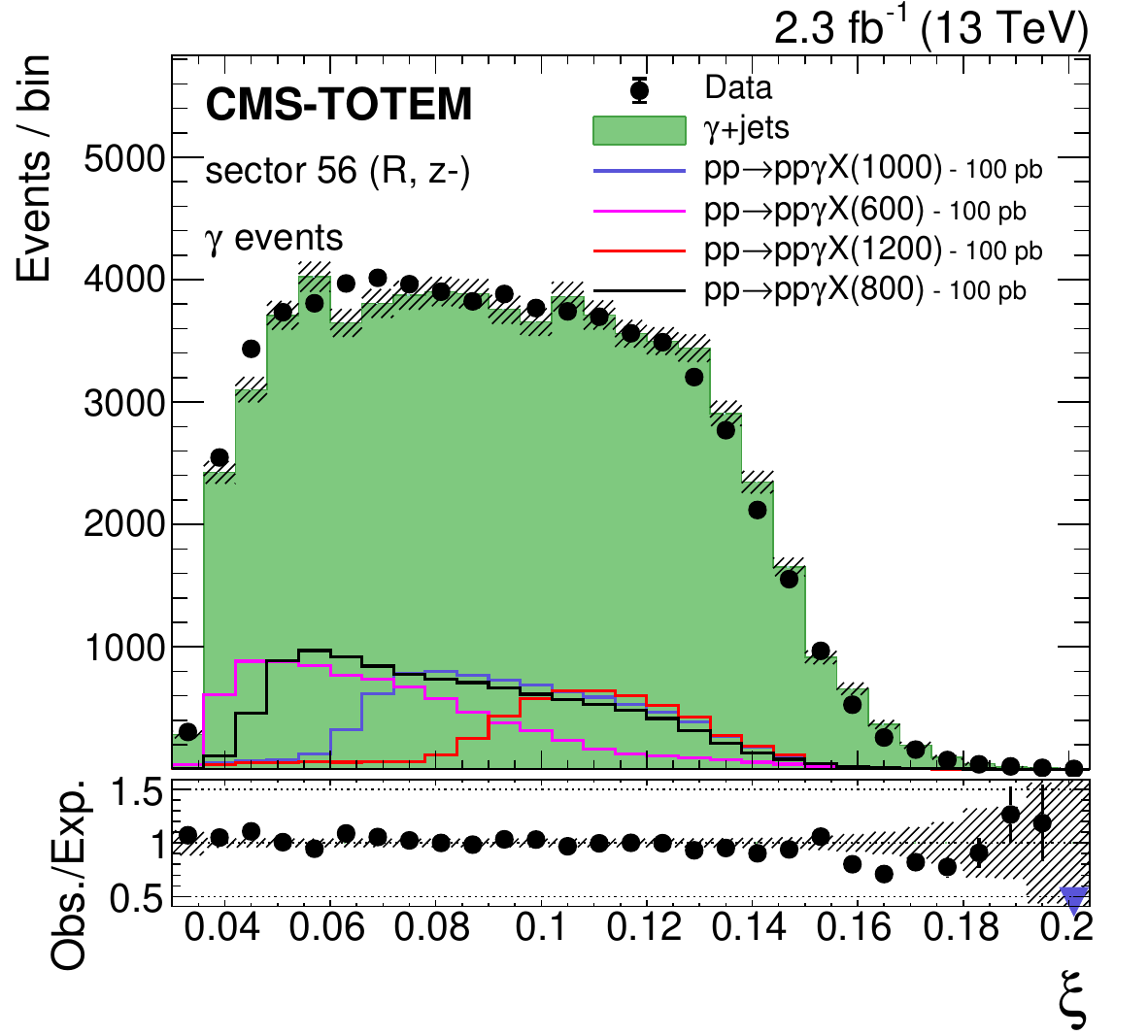}
\includegraphics[width=.32\textwidth]{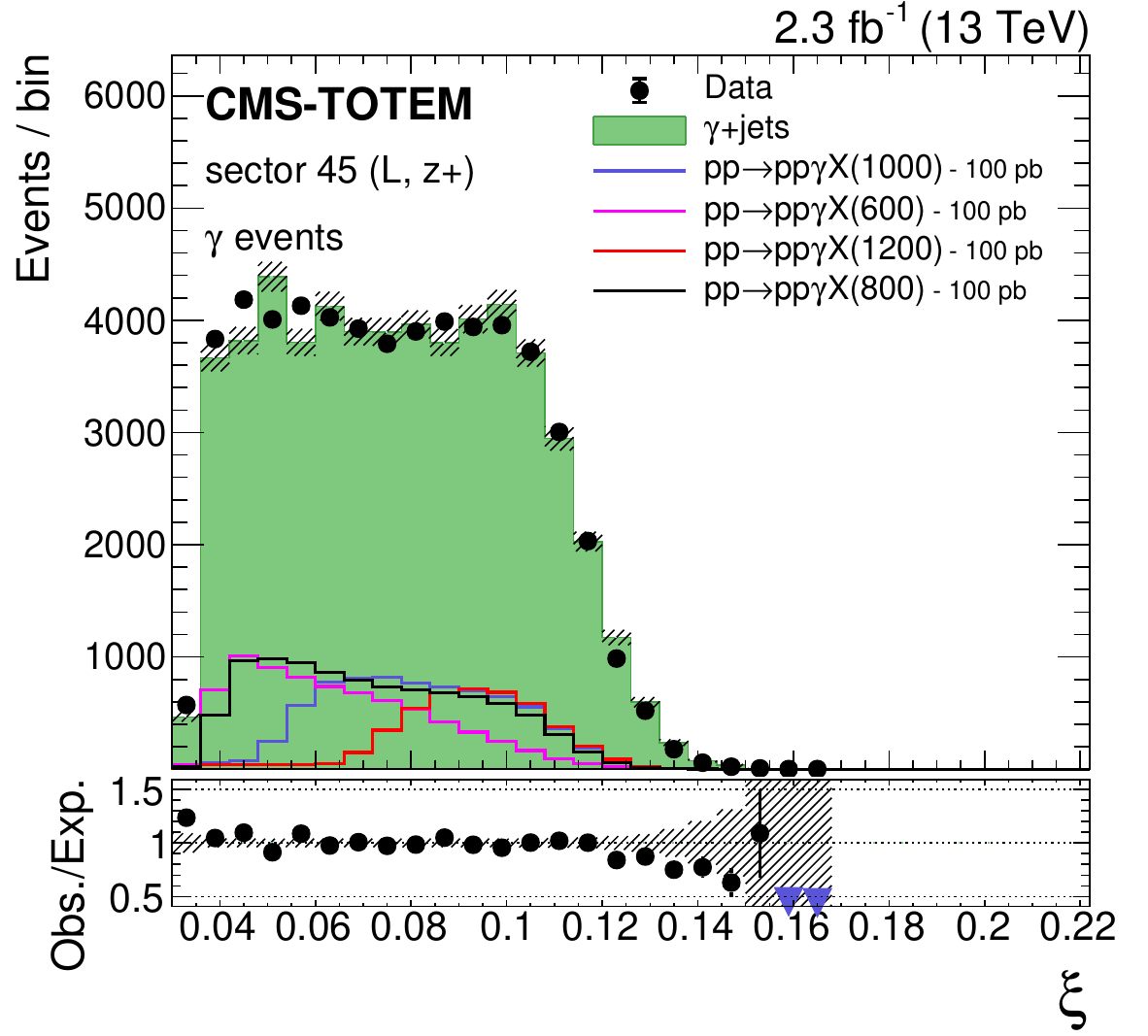}
\includegraphics[width=.32\textwidth]{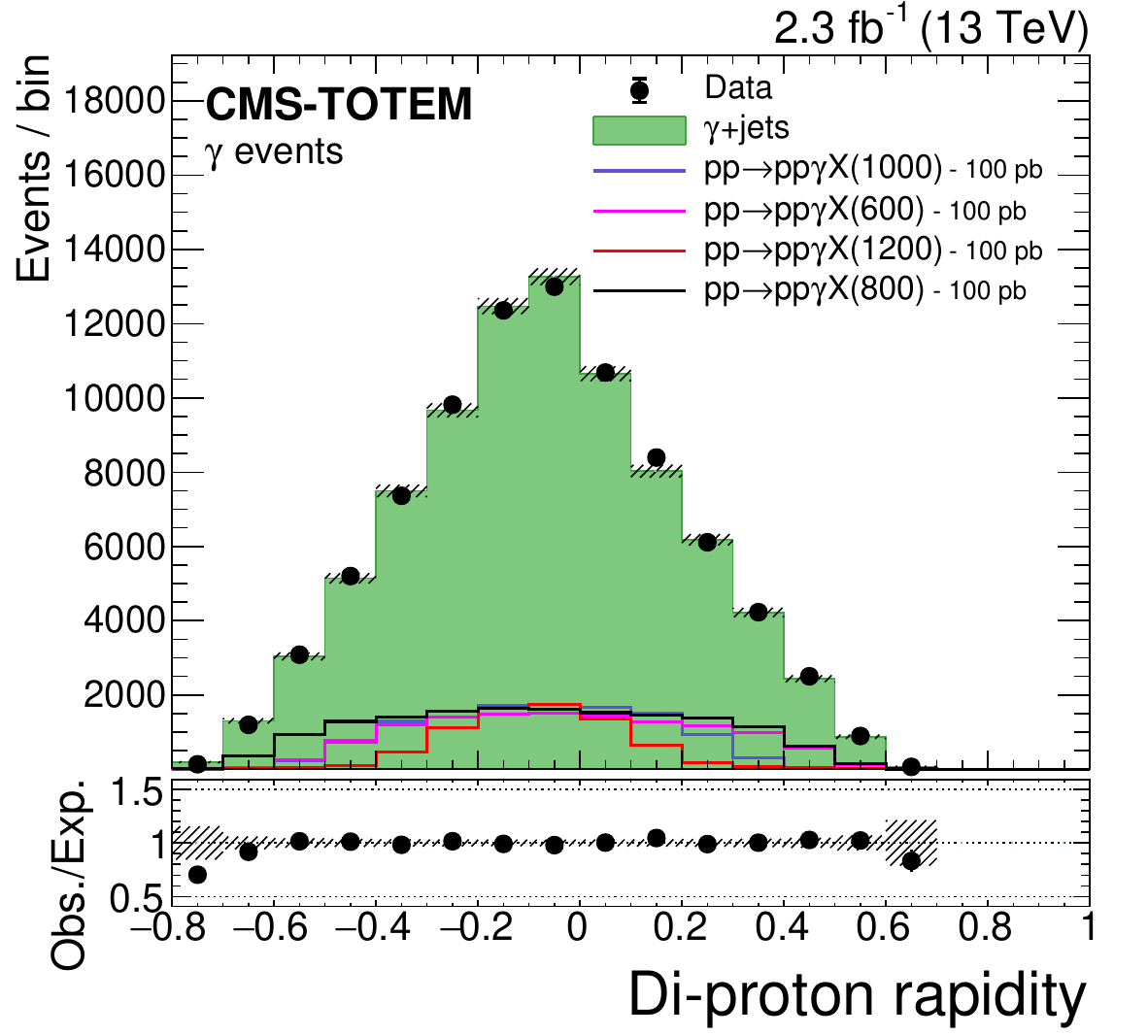}
\caption{
Distributions of the reconstructed proton $\xi$ in the negative arm (left), positive arm (middle), and the corresponding di-proton rapidity (right) from the proton mixing procedure with simulated MC events are compared to data, in the upper panels in each plot.
Processes other than the ones displayed in the figures are estimated to have negligible residual contributions.
The lower panels display the ratio between the data and the background model, 
with the arrows indicating values lying outside the displayed range.
The hatched band illustrates the statistical uncertainty of the background model.
The $\Pe\Pe$, $\Pgm\Pgm$, and photon final states are shown from top to bottom.
The $\Pe\Pe$ and $\Pgm\Pgm$ events are displayed without the \PZ boson \pt requirement. For illustration, the simulated signal distributions are superimposed for various choices of $\mX$, normalised to a generated fiducial cross section of 100\unit{pb}.
}
\label{fig:mccontrol}
\end{figure*}

A twofold validation of the background estimation method is performed.
A first check is made using MC simulation samples with embedded PU protons.
Figure~\ref{fig:mccontrol} illustrates the distributions of the reconstructed $\xi$ and di-proton rapidity for simulated events, compared to these two distributions seen in data.
In general, good agreement is observed, indicating that the samples are dominated by
inclusive production of \PZ and \PGg bosons with two forward protons from PU.

\begin{figure*}
\centering
\includegraphics[width=.44\textwidth]{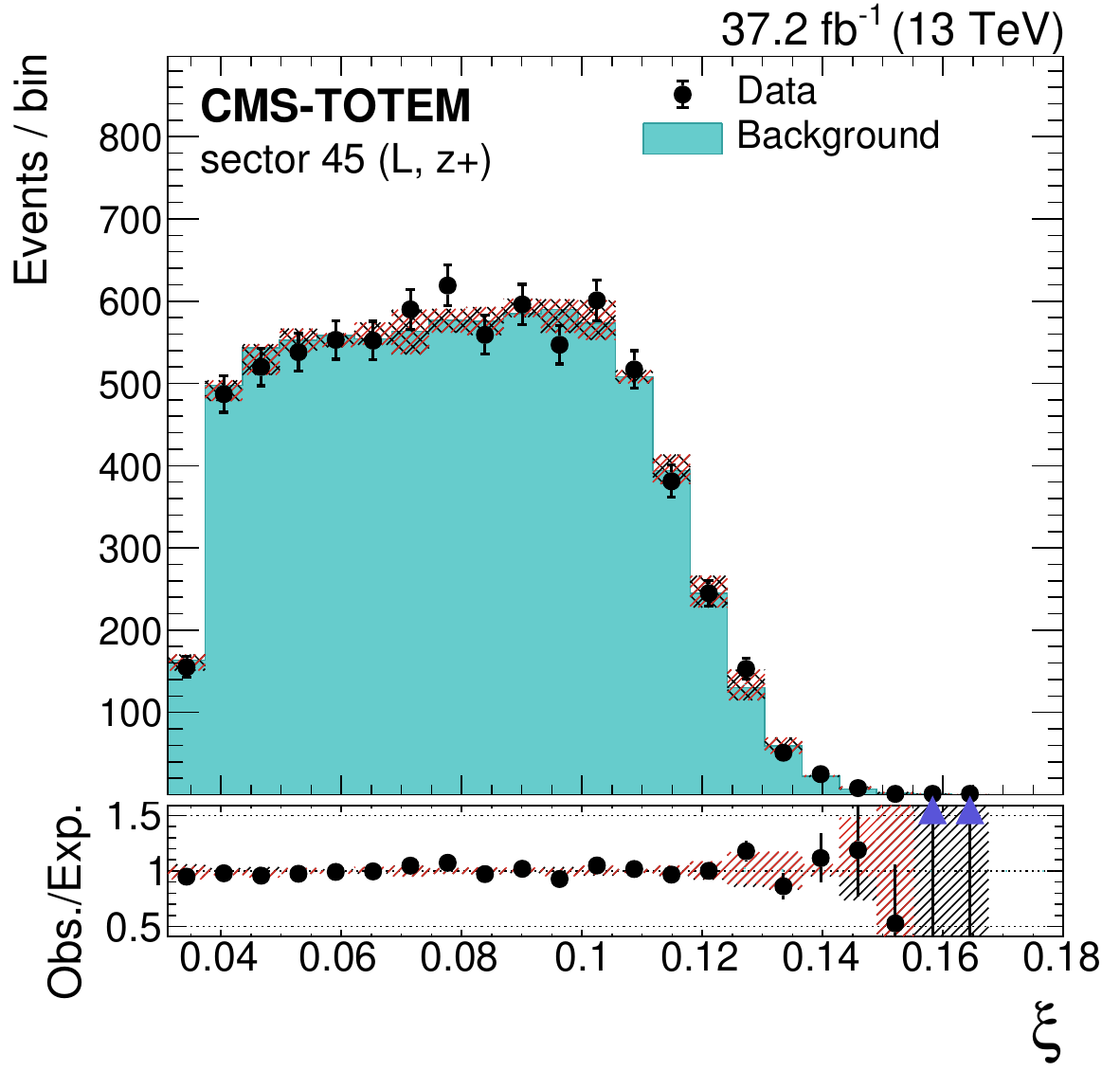}
\includegraphics[width=.44\textwidth]{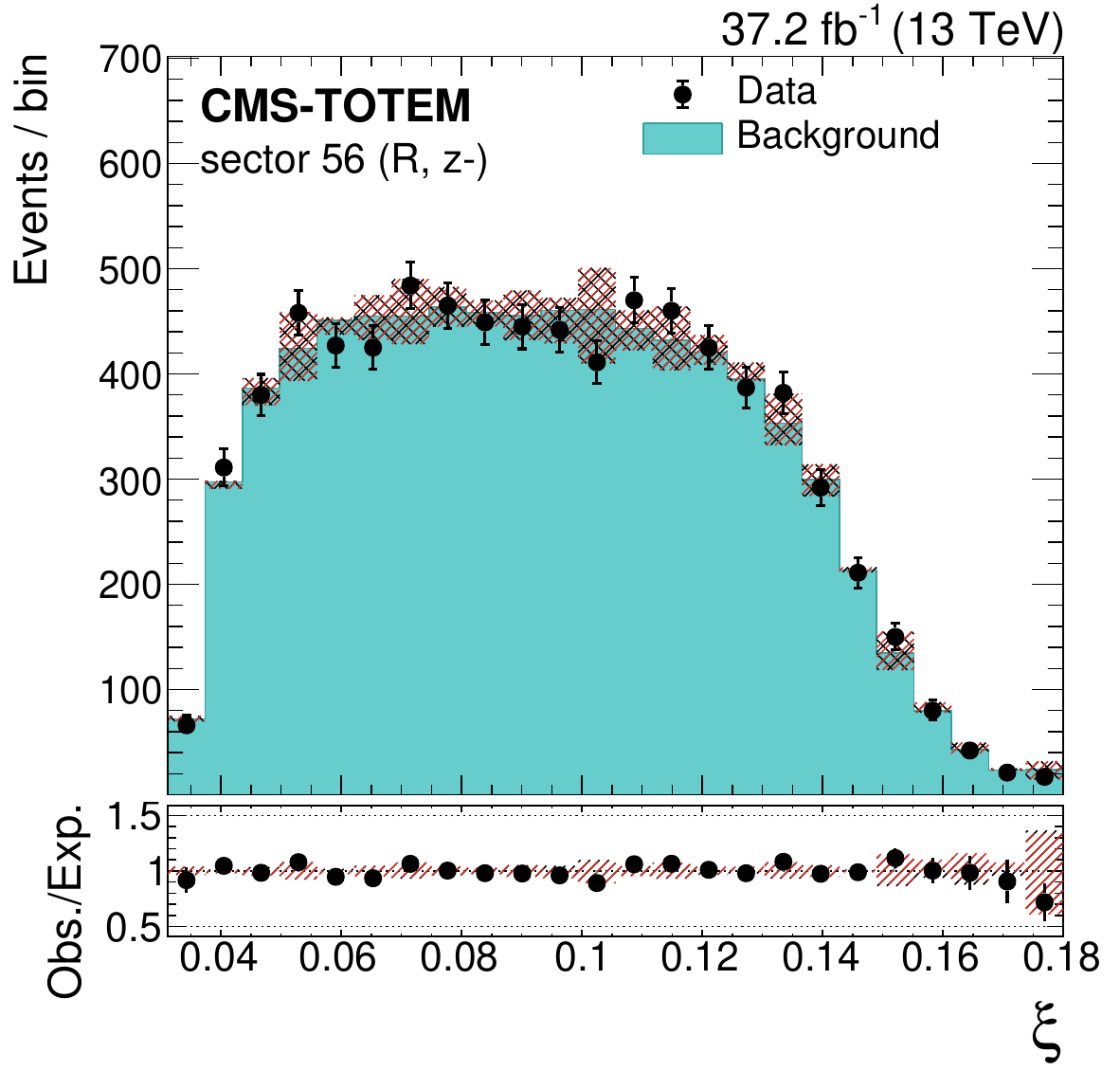}
\includegraphics[width=.44\textwidth]{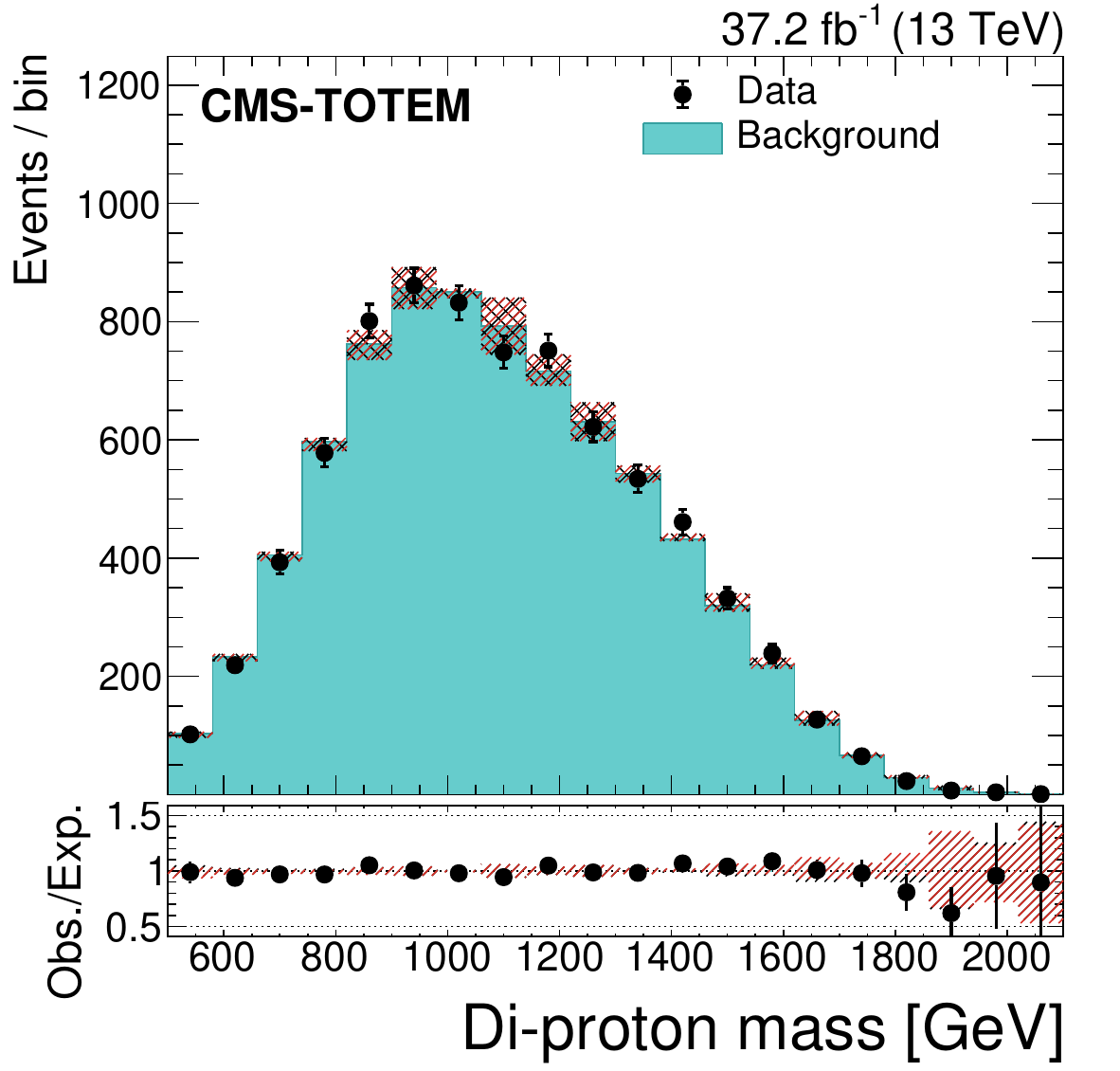}
\includegraphics[width=.44\textwidth]{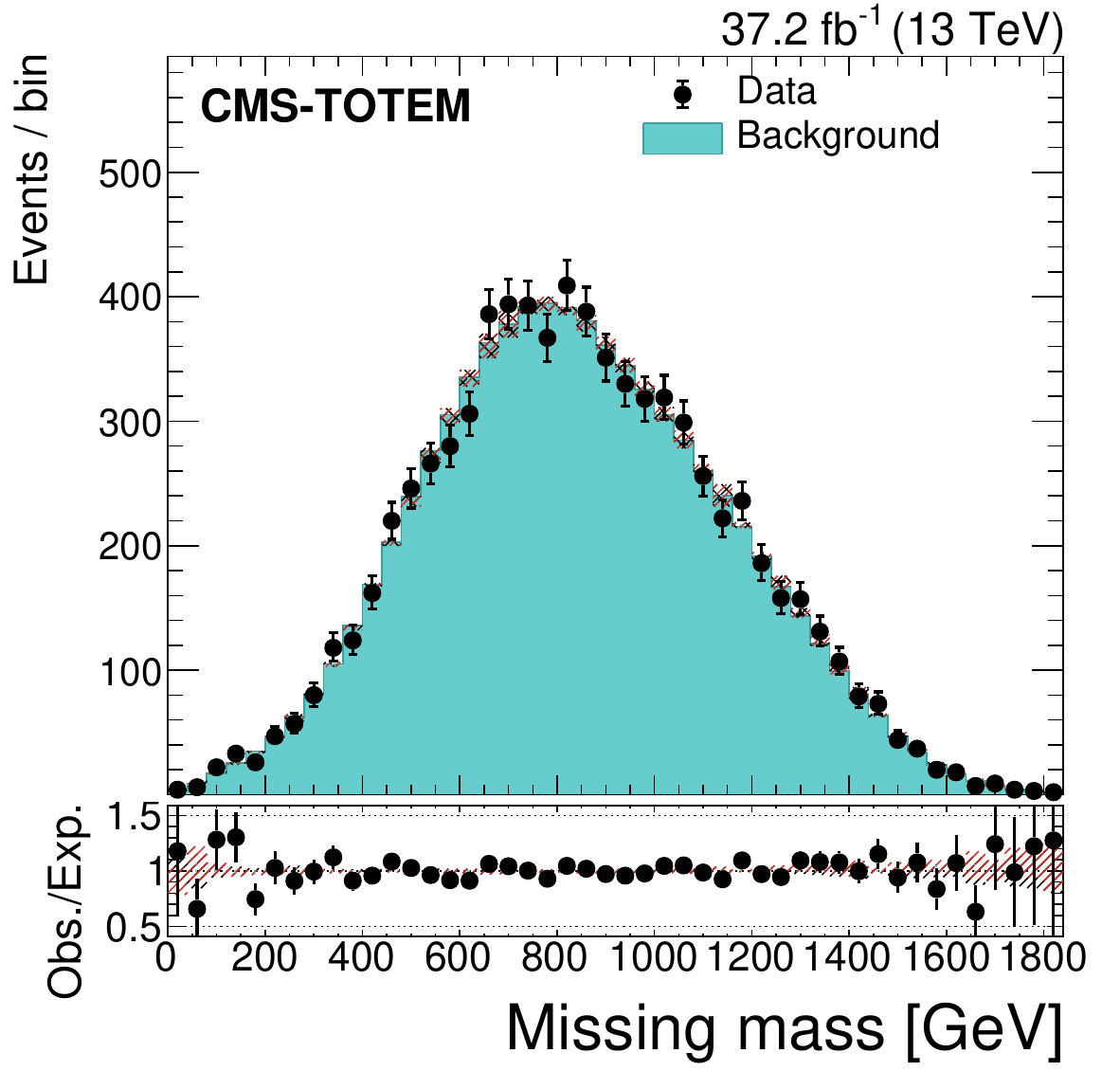}
\caption{Validation of the background modelling method, using the $\Pe\Pgm$ control sample. Selected $\Pe\Pgm$ events are mixed with protons from $\PZ\to\Pgm\Pgm$ events with $\pt(\PZ)<10\GeV$ to simulate the combinatorial background shape, while the data points are unaltered $\Pe\Pgm$ events. The proton $\xi$ distributions for both CT-PPS arms (upper row), those of the di-proton invariant mass (lower left), and of \mmiss (lower right) are shown. 
The lower panel in each plot displays the ratio between the data and the background model,
with the arrows indicating values lying outside the displayed range.
The gray uncertainty band around the background prediction represents the contribution from the limited sample size.
The red uncertainty band represents the effect of adding in quadrature the differences with the alternative mixing approaches described in the text. }
\label{fig:emcontrol}
\end{figure*}

A second check is made using the $\Pe\Pgm$ data control sample.
The $\Pe\Pgm$ data control sample is expected to be dominated by \ttbar events and to contain no signal events. Figure~\ref{fig:emcontrol} illustrates some of the distributions obtained with this sample. Selected $\Pe\Pgm$ events are mixed multiple times with protons from random $\PZ\to\Pgm\Pgm$ events with $\pt(\PZ)<10\GeV$ to simulate the combinatorial background shape, which is compared to the shape observed in unaltered $\Pe\Pgm$ events. 
The observed $\xi$ distributions, those of the di-proton invariant mass, and of \mmiss obtained with the combinatorial background model are in good agreement with the unaltered data.
The event selection is representative of the final event selection in the signal region,
with exactly one multi-RP proton candidate in each arm and a $\pt(\Pe\Pgm)>40\GeV$ requirement applied in the signal selection.
The latter is used as a requirement analogous to that on $\pt(\PZ)$.
The uncertainty band reflects the uncertainty from the 
the limited event sample size and the comparison with the background shapes
(either from mixing with protons from $\Pe\Pgm$ events or from mixing protons separately in each arm from $\PZ\to\Pgm\Pgm$).

\subsection{Proton mixing in simulated signal samples}
To model protons originating from PU interactions in the signal MC samples, an approach is used similar to that for the background modelling in data.
We add protons from events in the data control sample to simulated events, making sure that the protons are selected proportionally to the relative fractions of integrated luminosity for each of the data-taking periods. This procedure guarantees that the effects of PU protons on the missing mass reconstruction are correctly reproduced. As these protons are indistinguishable in the analysis from the signal protons, their presence can cause events outside the fiducial region to pass the event selection, or they may replace a signal proton in the \mmiss reconstruction, thus leading to a dilution of the signal, as is visible in Fig.~\ref{fig:mmass_signal_out}. The PU protons mixing procedure takes the strip inefficiency into account and propagates its effect to the multi-RP reconstruction efficiency.
However, when mixing protons from different events, an additional correction needs to be applied in the case no proton was reconstruced in the strip detectors. This could either have been caused by the absence of a proton within detector acceptance, or the presence of multiple protons. In the mixing procedure in signal samples, both options are taken into account with relative probabilities derived as a function of the run conditions, based on dedicated measurements using truly empty events.

\section{Statistical analysis and systematic uncertainties}
\label{sec:systuncs}

Using the signal and background \mmiss distributions,
we perform a statistical analysis to search for central exclusive $\PV\PX$ production.
The search is carried out for different categories (a total of 96 subsamples) of the data with different proton reconstruction algorithms, different beam crossing angles and different primary vertex multiplicities.
Depending on the proton reconstruction algorithm used, different binnings within the range 0--2\TeV are used. Bins with a width of 25\GeV are used when both protons are reconstructed with the multi-RP algorithm, 50\GeV when one of the protons is reconstructed with the pixel detector only, and 75\GeV when both protons are reconstructed with the pixels only. The choice of these values is related to the 
resolution of the reconstruction algorithms used in these categories.
While the bin width varies depending on the proton reconstruction algorithm used, it is independent of the mass in the region used for the search and reflects the CT-PPS resolution for \mmiss reconstruction.

A likelihood model containing three components is built to describe the \mmiss spectrum: 
a signal and two background contributions.
The background is composed of the combinatorial and the signal-induced contribution outside the fiducial volume.
The normalisations of each of these components are left to float freely in the fit,
with the normalisation of the contribution outside the fiducial volume independent of that of the signal, to reduce any dependence on the exact shape of the $p_z(\Pp\Pp)$ spectrum. The statistical analysis is based on \textsc{RooStats}~\cite{RooStats}.

The systematic uncertainties associated with the background and signal predictions are incorporated as nuisance parameters in the profile likelihood fit. They may affect the shapes of the expected \mmiss distributions of the signal and backgrounds, as well as their normalisations.
We consider the nuisance parameters uncorrelated between signal and background shapes, and between different categories,
even if similar methods are used to estimate the uncertainties.
The following systematic uncertainties are considered:

\begin{description}

\item[Pileup proton spectra:] The \mmiss shape for the background models is determined after replacing the protons of the selected events with protons from low-\pt{} $\PZ\to\Pgm\Pgm$ events.
For the signal the same source is used to model the contribution from PU protons.
As an alternative source of PU protons we use $\Pe\Pgm$ events. The difference between the two \mmiss shapes is determined point by point and symmetrically applied as a pre-fit estimate of the uncertainties associated with the PU proton spectra.
This systematic effect mostly affects the background shape, and the impact is approximately 4\% on average.
There is a significant difference in statistical precision between the two samples;
in some bins this generates large spurious variations (\eg in the tails of distributions) that are mitigated with a smoothing procedure.

\item[Single-diffractive backgrounds:] Given that we do not explicitly separate the background into its components (pure combinatorial, single diffractive, and double diffractive),
we add an uncertainty to estimate the possible difference in shape by mixing protons from a single arm instead of two.
The differences in shape between the positive-only (or negative-only) mixing and the double mixing are taken as an estimate of this uncertainty.
The impact on the background shape is approximately 2\% on average.

\item[CT-PPS efficiency:] The main source of uncertainty is the inefficiency induced by radiation damage in the strip detectors. The efficiency itself is taken into account by reweighting the
signal simulation. The weights are then varied according to their uncertainty, to generate alternative \mmiss shapes.
This uncertainty is in the range 2-5\%, depending on the event category.

\item[Time dependence:] The CT-PPS conditions were not uniform throughout the 2017 data taking.
This is taken into account by simulating independently each crossing angle and the two main data taking configurations before and after the so-called second technical stop (TS2) of the LHC~\cite{Todd:2294850}.
We assign a time-dependent uncertainty by varying the relative contribution of the pre- and post-TS2 simulations; this impacts the signal by approximately 1\% on average.
The weights are varied by 3\%, which reflects the overall uncertainty on the estimated integrated luminosity for the 2017 data set~\cite{CMS-PAS-LUM-17-004}.

\item[$p_{z}(\Pp\Pp)$ spectrum:] it is obtained with the equivalent photon approximation, as described in Section~\ref{subsec:signalmodel}, and its width is parametrised as function of $\mX$. We estimate the impact of the uncertainty of this parameterisation by varying the width at the generator level for each mass point (typically 10\%). The difference with respect to the nominal prediction is used as an uncertainty and is estimated to alter the signal shape by $<$1\%.

\item[Selection efficiency:] The uncertainties in the trigger and selection efficiencies for each final state have a negligible effect on the signal shape and only cause an approximately 3\% normalisation uncertainty.

\item[Integrated luminosity:] An additional 2.3\% normalisation uncertainty is assigned to the normalisation of the signal~\cite{CMS-PAS-LUM-17-004}.

\item [Limited event count:] Residual uncertainties due to limited sample sizes are also included in the fit using the Beeston--Barlow method~\cite{Barlow:1993dm}. They are typically small for the signal and background (${\leq}1\%$).

\end{description}

\section{Results}
\label{sec:results}

A total of 948~070 events with an $\Pe\Pe$ and 1~477~237 with a $\Pgm\Pgm$ final state, as well as 85~024 with a photon in the final state are selected for the analysis.
Events with dilepton final states ($\Pe\Pe$ and $\Pgm\Pgm$) originate almost exclusively
($>99.7\%$) from $\PZ\to\Pell\Pell$ production,
while the majority of those with a \PGg final state are estimated to come from single-photon production ($>99.8\%$).
The MC-based expectations agree with the data within 1--10\%, depending on the final state and the subsample.
In the statistical analysis we use the total number of events selected in the data as an initial estimate for the normalisation of the combinatorial background, so these residual normalisation differences have no impact on the results.

\begin{figure*}[!htp]
\centering
\includegraphics[width=.99\textwidth]{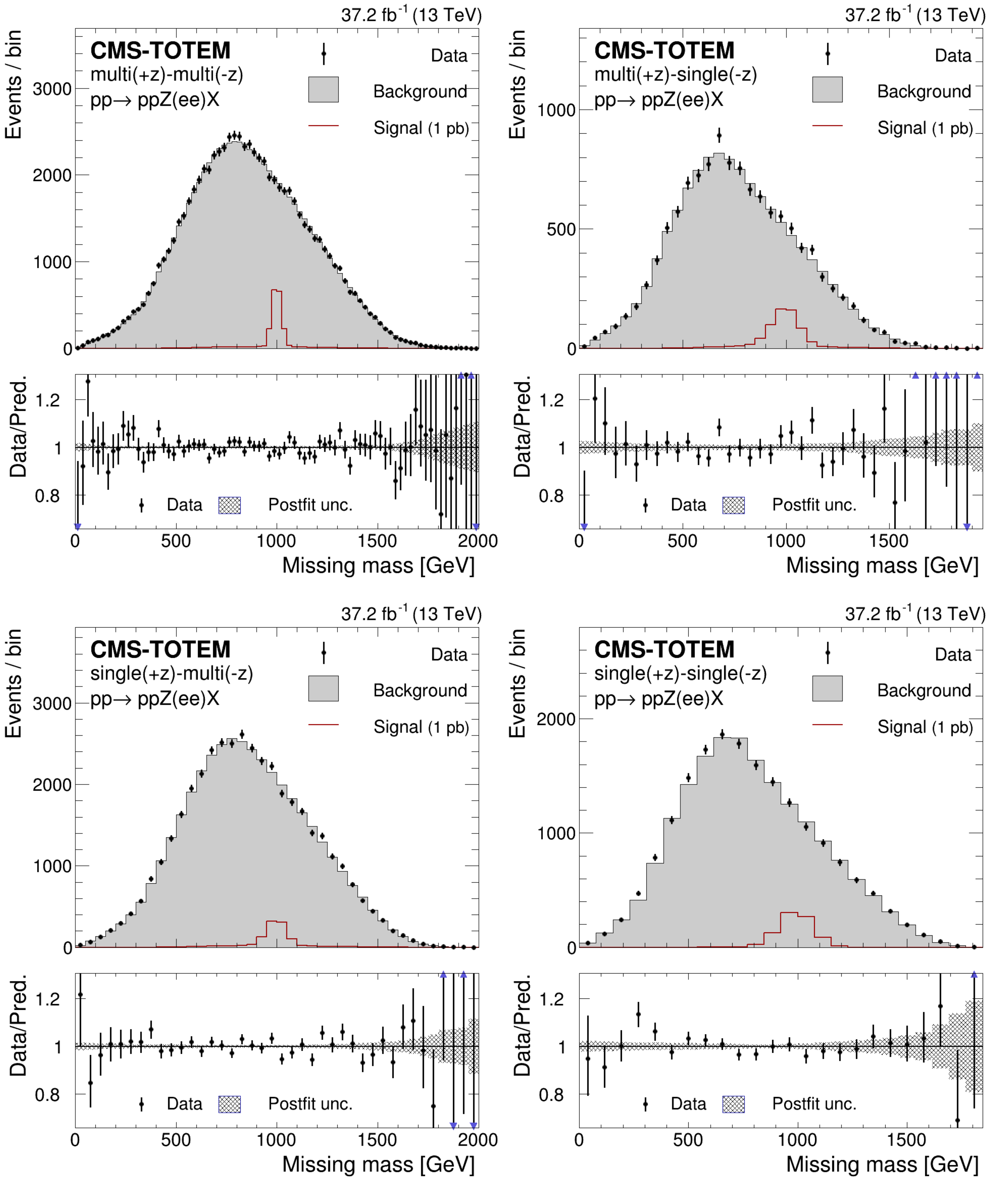}
\caption{
\mmiss distributions in the $\PZ\to \Pe\Pe$ final state.
The distributions are shown for protons reconstructed with (from left to right) the multi-multi, multi-single, single-multi and single-single methods, respectively.
The background distributions are shown after the fit.
The lower panels display the ratio between the data and the background model,
with the arrows indicating values lying outside the displayed range.
The expectations for a signal with  $\mX = 1000\GeV$ are superimposed and normalised to 1\unit{pb}.
}
\label{fig:mmass_final_zee}
\end{figure*}

\begin{figure*}[!htp]
\centering
\includegraphics[width=.99\textwidth]{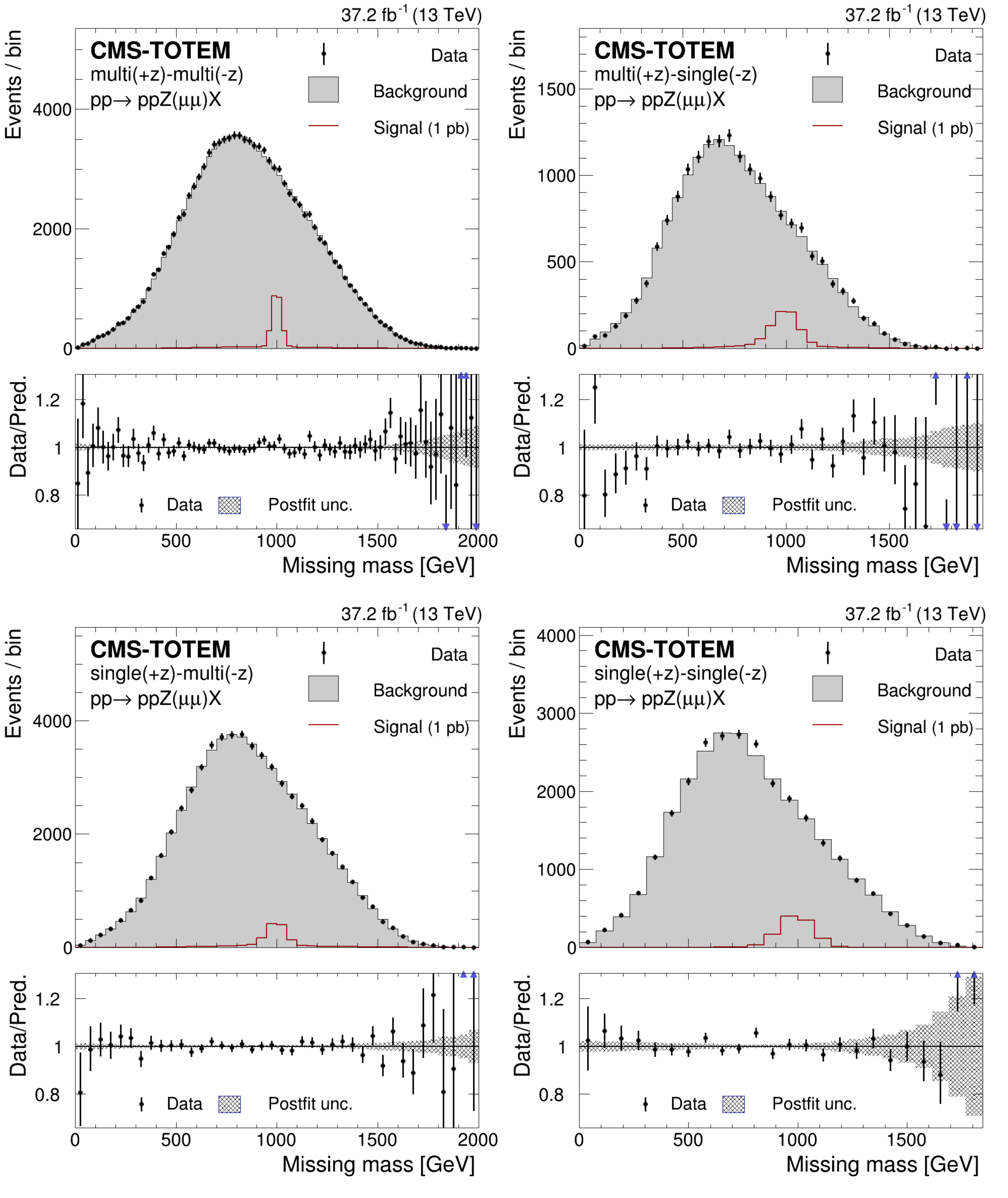}
\caption{
\mmiss distributions in the $\PZ\to\Pgm\Pgm$ final state.
The distributions are shown for protons reconstructed with (from left to right) the multi-multi, multi-single, single-multi and single-single methods, respectively.
The background distributions are shown after the fit.
The lower panels display the ratio between the data and the background model,
with the arrows indicating values lying outside the displayed range.
The expectations for a signal with  $\mX = 1000\GeV$ are superimposed and normalised to 1\unit{pb}.
}
\label{fig:mmass_final_zmm}
\end{figure*}

\begin{figure*}[!htp]
\centering
\includegraphics[width=.99\textwidth]{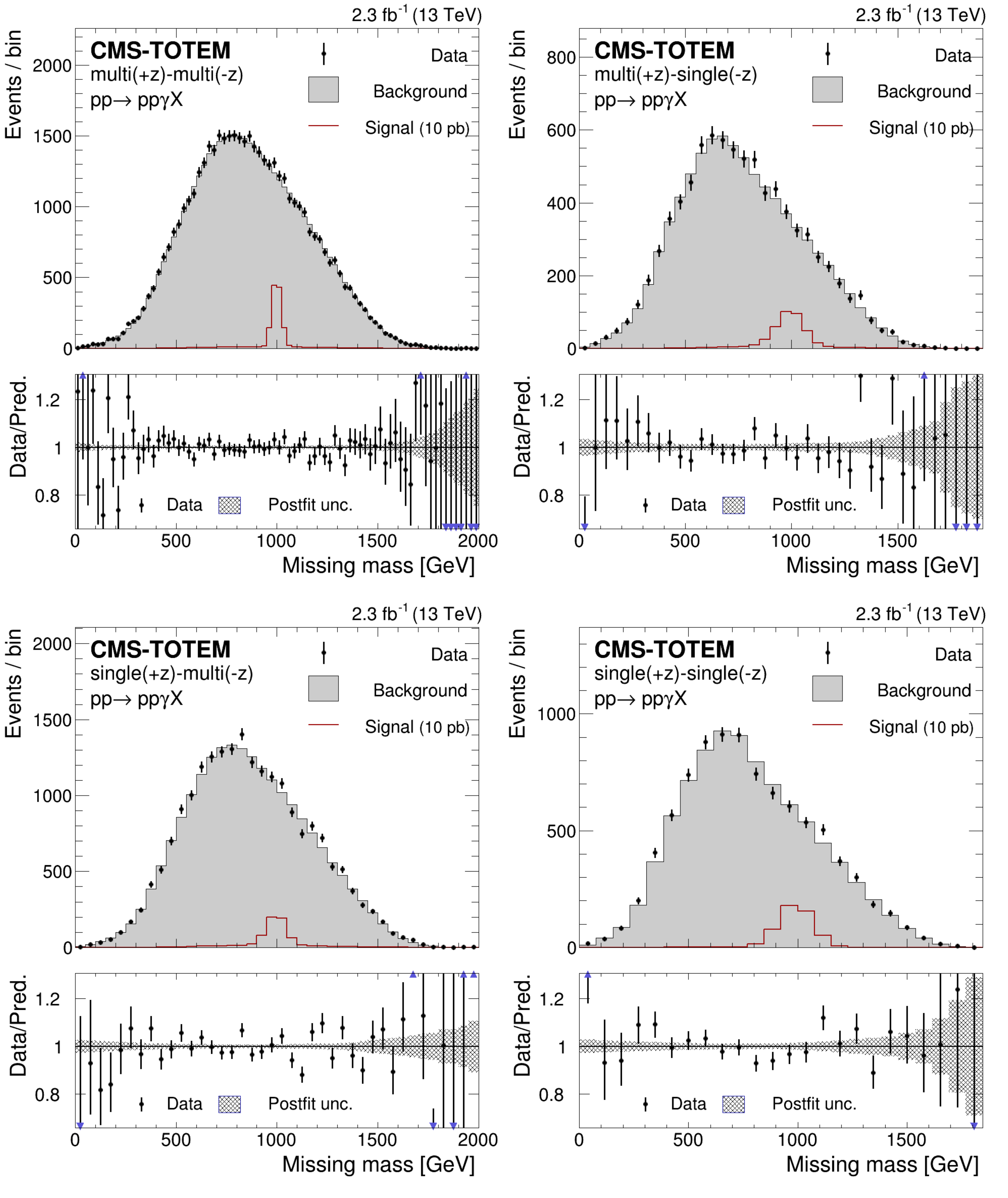}
\caption{
\mmiss distributions in the \PGg final state.
The distributions are shown for protons reconstructed with (from left to right) the multi-multi, multi-single, single-multi and single-single methods, respectively.
The background distributions are shown after the fit.
The lower panels display the ratio between the data and the background model, with the arrows indicating values lying outside the displayed range.
The expectations for a signal with  $\mX = 1000\GeV$ are superimposed and normalised to 10\unit{pb}.}
\label{fig:mmass_final_a}
\end{figure*}

The observed distributions of the \mmiss variable are shown in Figs.~\ref{fig:mmass_final_zee}, \ref{fig:mmass_final_zmm}, and \ref{fig:mmass_final_a}.
The distributions are plotted after the fit is performed and therefore include the adjustments to the background shape and
normalisation derived from the observed data. The normalisation of the out-fiducial component after the fit is typically close to 0.
A reference signal is overlaid for illustration. Data and background expectation agree within 10\% or better in all subsamples.

\begin{figure*}[!htp]
\centering
\includegraphics[width=.49\textwidth]{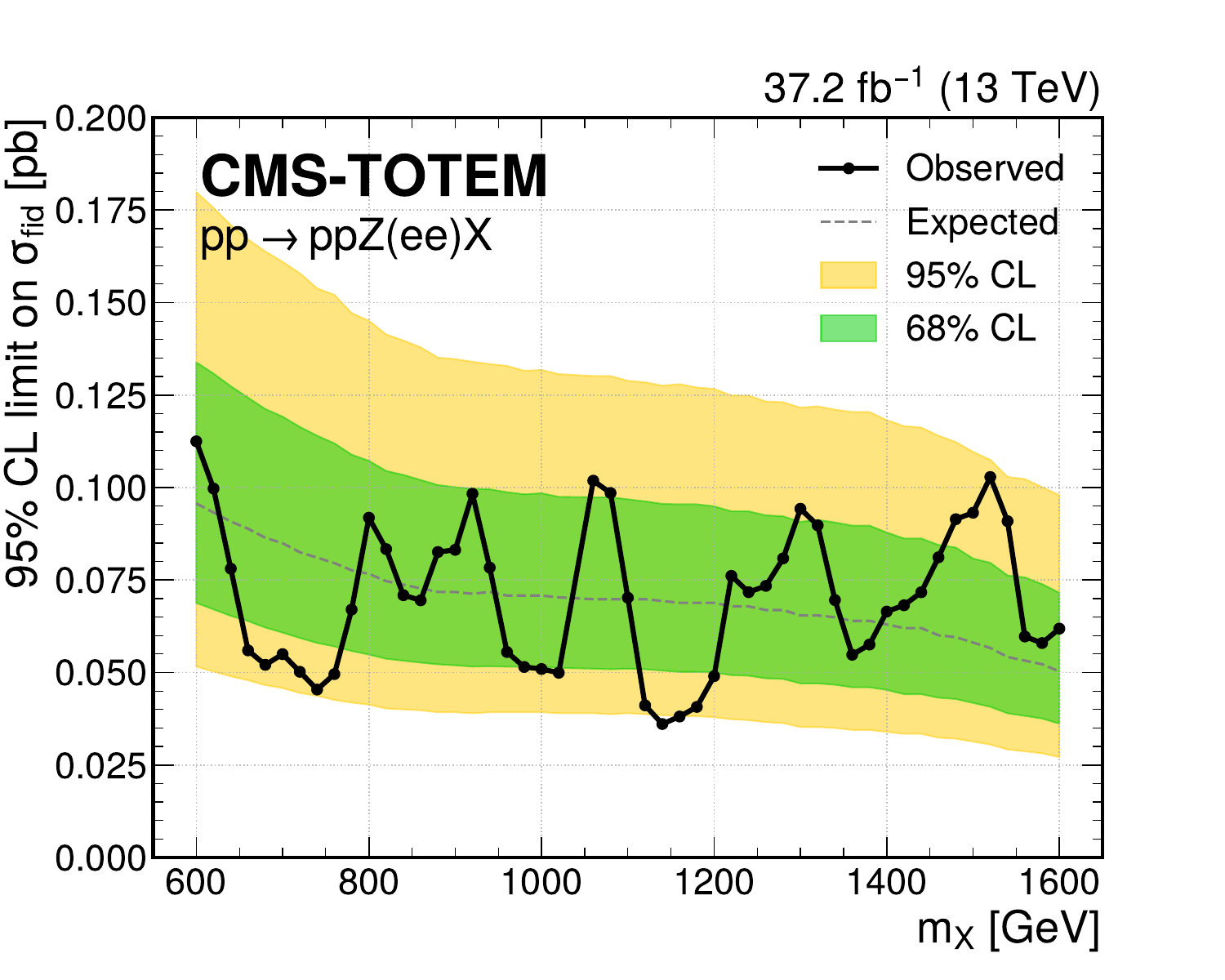}
\includegraphics[width=.49\textwidth]{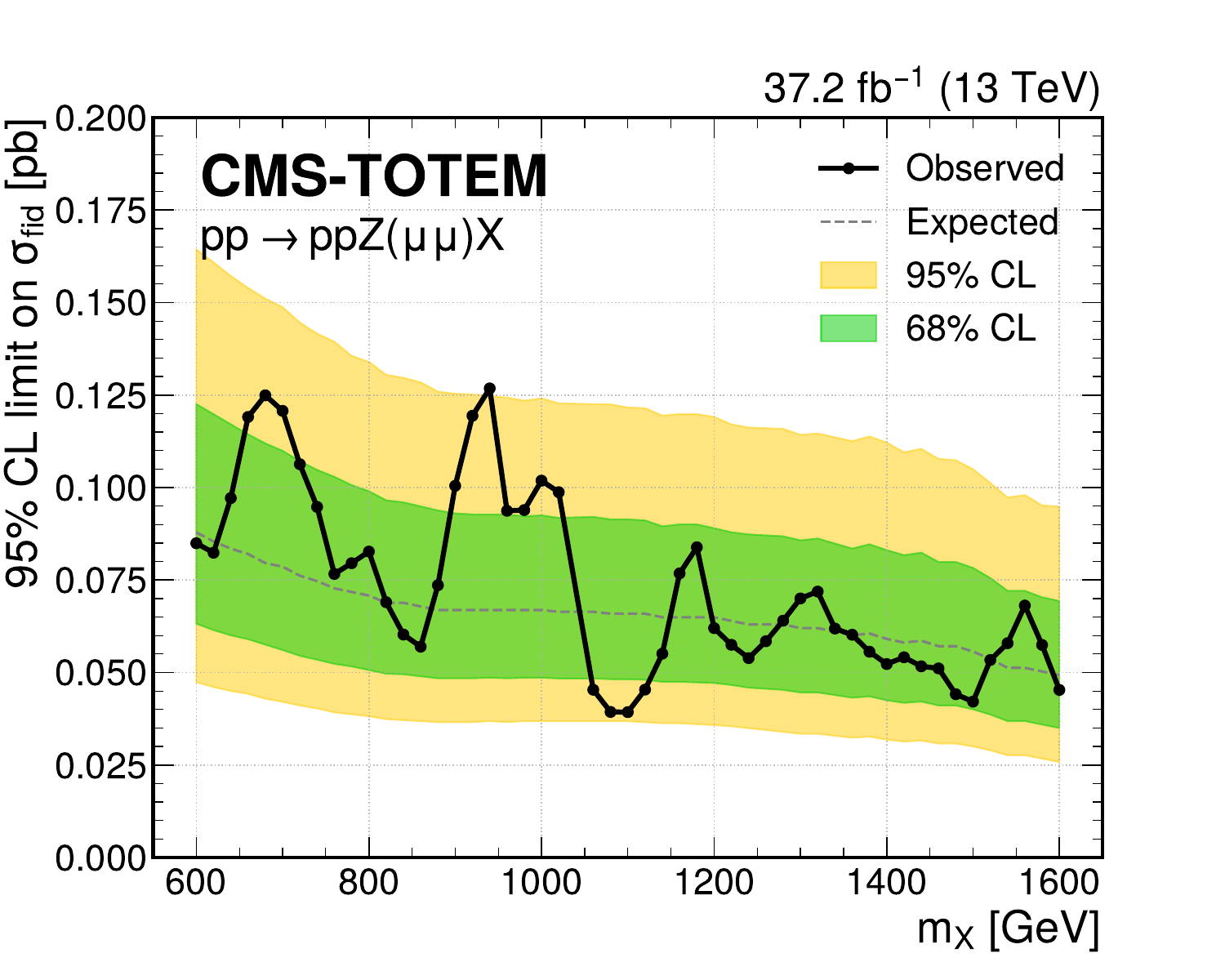}
\includegraphics[width=.49\textwidth]{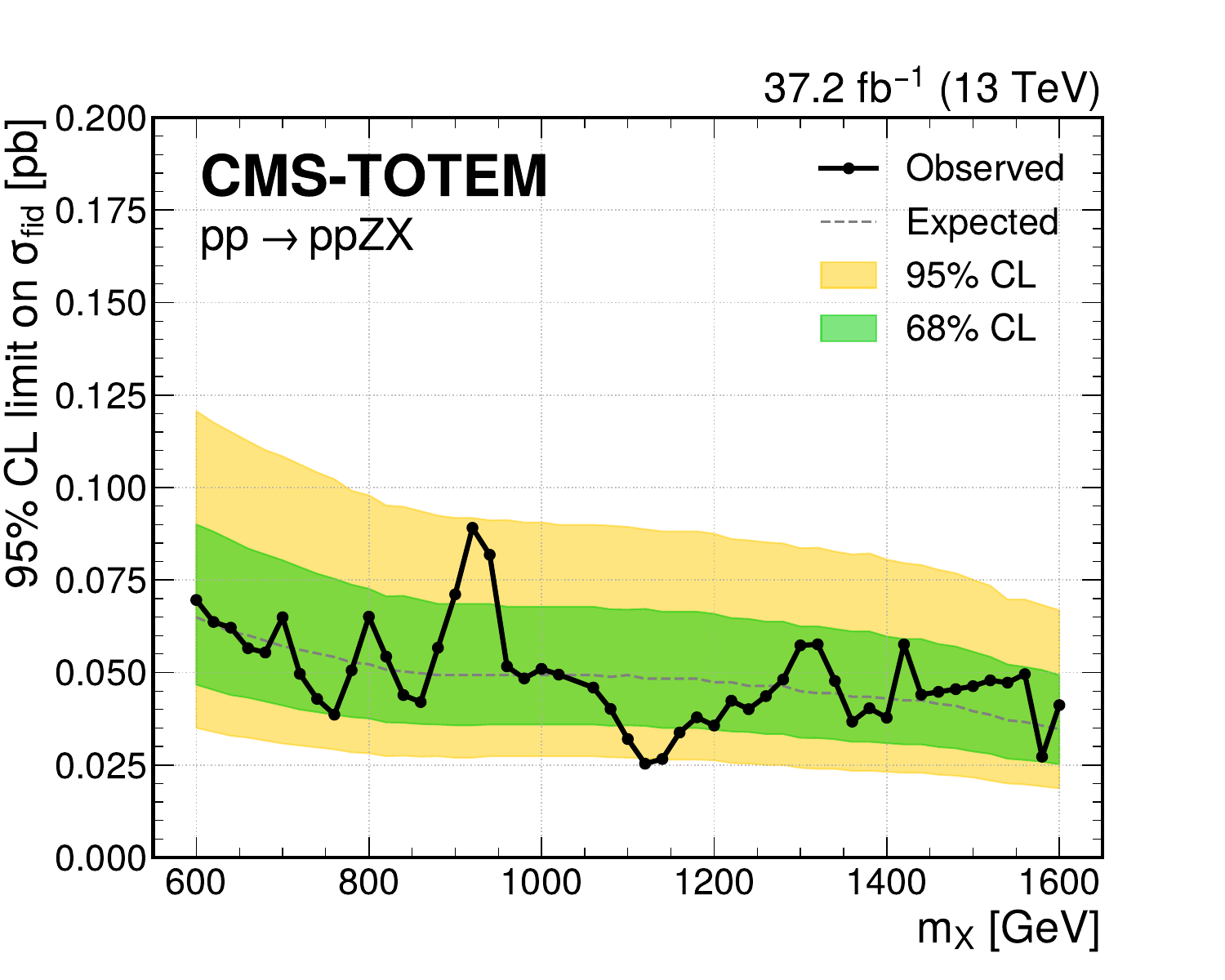}
\includegraphics[width=.49\textwidth]{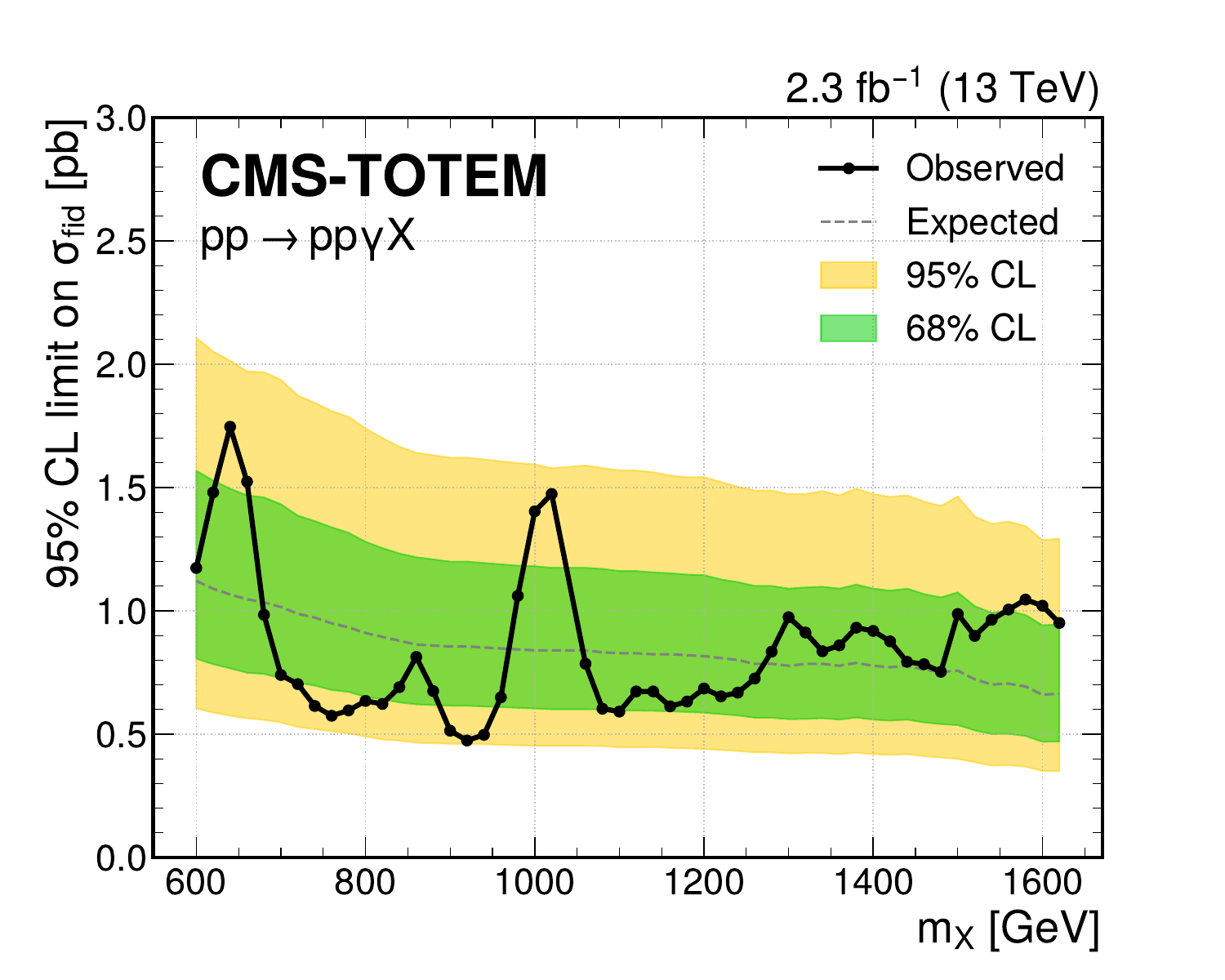}
\caption{
Upper limits on the $\Pp\Pp\to \Pp\Pp\PZ/\PGg+\PX$ cross section at 95\% \CL, as a function of $\mX$.
The 68 and 95\% central intervals of the expected limits are represented by the dark green and light yellow bands, respectively,
while the observed limit is superimposed as a curve.
The upper plots correspond to the $\PZ\to \Pe\Pe$ and $\PZ\to\Pgm\Pgm$ final states,
while the lower plots correspond to the combined \PZ and \PGg analyses.
}
\label{fig:limits}
\end{figure*}

We set limits on the cross section for the anomalous central exclusive reaction ${\Pp\Pp\to \Pp\Pp\PZ/\PGg+\PX}$.
The 95\% confidence level (\CL) upper limits are calculated using a modified frequentist approach with the \CLs criterion~\cite{Junk:1999kv,Read_2002}.
An asymptotic approximation is used for the test statistic~\cite{ATLAS:2011tau,Cowan:2010js}.
The limits obtained are shown in Fig.~\ref{fig:limits}.
Overall the observed limits are within the 95\% central interval 
of the expected limits, within the explored mass range.
The fluctuations in the observed limit are compatible with expectations from effects due to the detector resolution.
The limits translate into local $p$-values~\cite{lee} supporting
the background-only hypothesis within two standard deviations.

\ifthenelse{\boolean{cms@external}}{\clearpage}{}
\section{Summary}
\label{sec:conclusions}

A search is presented for anomalous central exclusive $\PZ/\PGg+\PX$ production using proton-proton ($\Pp\Pp$) data samples corresponding 
to an integrated luminosity up to $37.2\fbinv$ recorded in 2017 by the CMS detector and the CMS-TOTEM precision proton spectrometer (CT-PPS). A hypothetical \PX resonance is searched for in the mass region between 0.6 and 1.6\TeV, with selections optimised for the best expected significance.
Benefitting from the excellent mass resolution of 2\% offered by the combination of the CMS central detector and CT-PPS, for the first time at the CERN Large Hadron Collider (LHC), the missing mass distribution is used to perform a model-independent search. Upper limits on the visible cross section of the $\Pp\Pp\to \Pp\Pp \PZ/\PGg+\PX$ process are set in a fiducial volume, using a generic model, in the absence of significant deviations in data with respect to the background predictions.
Upper limits in the 0.025--0.089\unit{pb} range are obtained for $\sigma(\Pp\Pp\to \Pp\Pp\PZ\PX)$ and 0.47--1.75\unit{pb} for $\sigma(\Pp\Pp\to \Pp\Pp\PGg\PX)$.
With these results we demonstrate the feasibility of the missing mass approach for searches at the LHC.

\begin{acknowledgments}
We congratulate our colleagues in the CERN accelerator departments for the excellent performance of the LHC and thank the technical and administrative staffs at CERN and at other CMS and TOTEM institutes for their contributions to the success of the common CMS-TOTEM effort. In addition, we gratefully acknowledge the computing centers and personnel of the Worldwide LHC Computing Grid and other centers for delivering so effectively the computing infrastructure essential to our analyses. Finally, we acknowledge the enduring support for the construction and operation of the LHC, the CMS and TOTEM detectors, and the supporting computing infrastructure provided by the following funding agencies: BMBWF and FWF (Austria); FNRS and FWO (Belgium); CNPq, CAPES, FAPERJ, FAPERGS, and FAPESP (Brazil); MES and BNSF (Bulgaria); CERN; CAS, MoST, and NSFC (China); MINCIENCIAS (Colombia); MSES and CSF (Croatia); RIF (Cyprus); SENESCYT (Ecuador); MoER, ERC PUT and ERDF (Estonia); Academy of Finland, Magnus Ehrnrooth Foundation, MEC, HIP, and Waldemar von Frenckell Foundation (Finland); CEA and CNRS/IN2P3 (France); BMBF, DFG, and HGF (Germany); GSRI (Greece);  NKFIH (Hungary);  DAE and DST (India); IPM (Iran); SFI (Ireland); INFN (Italy); MSIP and NRF (Republic of Korea); MES (Latvia); LAS (Lithuania); MOE and UM (Malaysia); BUAP, CINVESTAV, CONACYT, LNS, SEP, and UASLP-FAI (Mexico); MOS (Montenegro); MBIE (New Zealand); PAEC (Pakistan); MES and NSC (Poland); FCT (Portugal); MESTD (Serbia); MCIN/AEI and PCTI (Spain); MOSTR (Sri Lanka); Swiss Funding Agencies (Switzerland); MST (Taipei); MHESI and NSTDA (Thailand); TUBITAK and TENMAK (Turkey); NASU (Ukraine); STFC (United Kingdom); DOE and NSF (USA).
        
\hyphenation{Rachada-pisek} Individuals have received support from the Marie-Curie program and the European Research Council and Horizon 2020 Grant, contract Nos.\ 675440, 724704, 752730, 758316, 765710, 824093, 884104, and COST Action CA16108 (European Union); the Leventis Foundation; the Alfred P.\ Sloan Foundation; the Alexander von Humboldt Foundation; the Belgian Federal Science Policy Office; the Fonds pour la Formation \`a la Recherche dans l'Industrie et dans l'Agriculture (FRIA-Belgium); the Agentschap voor Innovatie door Wetenschap en Technologie (IWT-Belgium); the F.R.S.-FNRS and FWO (Belgium) under the ``Excellence of Science -- EOS" -- be.h project n.\ 30820817; the Beijing Municipal Science \& Technology Commission, No. Z191100007219010; the Ministry of Education, Youth and Sports (MEYS) of the Czech Republic; Svenska Kulturfonden (Finland); the Hellenic Foundation for Research and Innovation (HFRI), Project Number 2288 (Greece); the Deutsche Forschungsgemeinschaft (DFG), under Germany's Excellence Strategy -- EXC 2121 ``Quantum Universe" -- 390833306, and under project number 400140256 - GRK2497; the Hungarian Academy of Sciences, the New National Excellence Program - \'UNKP, the NKFIH research grants K 124845, K 124850, K 128713, K 128786, K 129058, K 131991, K 133046, K 138136, K 143460, K 143477, 2020-2.2.1-ED-2021-00181, and TKP2021-NKTA-64 (Hungary); the Council of Science and Industrial Research, India; the Latvian Council of Science; the Ministry of Education and Science, project no. 2022/WK/14, and the National Science Center, contracts Opus 2021/41/B/ST2/01369 and 2021/43/B/ST2/01552 (Poland); the Funda\c{c}\~ao para a Ci\^encia e a Tecnologia, grant CEECIND/01334/2018 (Portugal); MCIN/AEI/10.13039/501100011033, ERDF ``a way of making Europe", and the Programa Estatal de Fomento de la Investigaci{\'o}n Cient{\'i}fica y T{\'e}cnica de Excelencia Mar\'{\i}a de Maeztu, grant MDM-2017-0765 and Programa Severo Ochoa del Principado de Asturias (Spain); the Chulalongkorn Academic into Its 2nd Century Project Advancement Project, and the National Science, Research and Innovation Fund via the Program Management Unit for Human Resources \& Institutional Development, Research and Innovation, grant B05F650021 (Thailand); the Kavli Foundation; the Nvidia Corporation; the SuperMicro Corporation; the Welch Foundation, contract C-1845; and the Weston Havens Foundation (USA).  
\end{acknowledgments}

\bibliography{auto_generated}
\cleardoublepage \appendix\section{The CMS Collaboration \label{app:collab}}\begin{sloppypar}\hyphenpenalty=5000\widowpenalty=500\clubpenalty=5000
\cmsinstitute{Yerevan Physics Institute, Yerevan, Armenia}
{\tolerance=6000
A.~Tumasyan\cmsAuthorMark{1}\cmsorcid{0009-0000-0684-6742}
\par}
\cmsinstitute{Institut f\"{u}r Hochenergiephysik, Vienna, Austria}
{\tolerance=6000
W.~Adam\cmsorcid{0000-0001-9099-4341}, J.W.~Andrejkovic, T.~Bergauer\cmsorcid{0000-0002-5786-0293}, S.~Chatterjee\cmsorcid{0000-0003-2660-0349}, K.~Damanakis\cmsorcid{0000-0001-5389-2872}, M.~Dragicevic\cmsorcid{0000-0003-1967-6783}, A.~Escalante~Del~Valle\cmsorcid{0000-0002-9702-6359}, P.S.~Hussain\cmsorcid{0000-0002-4825-5278}, M.~Jeitler\cmsAuthorMark{2}\cmsorcid{0000-0002-5141-9560}, N.~Krammer\cmsorcid{0000-0002-0548-0985}, L.~Lechner\cmsorcid{0000-0002-3065-1141}, D.~Liko\cmsorcid{0000-0002-3380-473X}, I.~Mikulec\cmsorcid{0000-0003-0385-2746}, P.~Paulitsch, F.M.~Pitters, J.~Schieck\cmsAuthorMark{2}\cmsorcid{0000-0002-1058-8093}, R.~Sch\"{o}fbeck\cmsorcid{0000-0002-2332-8784}, D.~Schwarz\cmsorcid{0000-0002-3821-7331}, S.~Templ\cmsorcid{0000-0003-3137-5692}, W.~Waltenberger\cmsorcid{0000-0002-6215-7228}, C.-E.~Wulz\cmsAuthorMark{2}\cmsorcid{0000-0001-9226-5812}
% ...
\par}
\vskip\cmsinstskip
\dag:~Deceased\\
$^{1}$Also at Yerevan State University, Yerevan, Armenia\\
$^{2}$Also at TU Wien, Vienna, Austria\\
% ...
\section{The TOTEM Collaboration\label{app:totem}}
\newcommand{\AddAuthor}[2]{#1$^{#2}$,\ }
\newcommand\AddAuthorLast[2]{#1$^{#2}$}
\noindent
\AddAuthor{G.~Antchev\cmsorcid{0000-0003-3210-5037}}{a}
\AddAuthor{P.~Aspell}{8}
\AddAuthor{I.~Atanassov\cmsorcid{0000-0002-5728-9103}}{a}
\AddAuthor{V.~Avati}{7,8}
\AddAuthor{J.~Baechler}{8}
\AddAuthor{C.~Baldenegro~Barrera\cmsorcid{0000-0002-6033-8885}}{10}
\AddAuthor{V.~Berardi\cmsorcid{0000-0002-8387-4568}}{4a,4b}
\AddAuthor{M.~Berretti\cmsorcid{0000-0003-4122-8282}}{2a}
\AddAuthor{V.~Borshch\cmsorcid{0000-0002-5479-1982}}{11}
\AddAuthor{E.~Bossini\cmsorcid{0000-0002-2303-2588}}{6a}
\AddAuthor{U.~Bottigli\cmsorcid{0000-0002-0666-3433}}{6c}
\AddAuthor{M.~Bozzo\cmsorcid{0000-0002-1715-0457}}{5a,5b}
\AddAuthor{H.~Burkhardt}{8}
\AddAuthor{F.S.~Cafagna\cmsorcid{0000-0002-7450-4784}}{4a}
\AddAuthor{M.G.~Catanesi\cmsorcid{0000-0002-2987-7688}}{4a}
\AddAuthor{M.~Csan\'{a}d\cmsorcid{0000-0002-3154-6925}}{3a,b}
\AddAuthor{T.~Cs\"{o}rg\H{o}\cmsorcid{0000-0002-9110-9663}}{3a,3b}
\AddAuthor{M.~Deile\cmsorcid{0000-0001-5085-7270}}{8}
\AddAuthor{F.~De~Leonardis}{4a,4c}
\AddAuthor{M.~Doubek\cmsorcid{0000-0002-0026-9558}}{1c}
\AddAuthor{D.~Druzhkin}{8,11}
\AddAuthor{K.~Eggert}{9}
\AddAuthor{V.~Eremin}{e}
\AddAuthor{A.~Fiergolski}{8}
\AddAuthor{F.~Garcia\cmsorcid{0000-0002-4023-7964}}{2a}
\AddAuthor{V.~Georgiev}{1a}
\AddAuthor{S.~Giani}{8}
\AddAuthor{L.~Grzanka\cmsorcid{0000-0002-3599-854X}}{7}
\AddAuthor{J.~Hammerbauer}{1a}
\AddAuthor{T.~Isidori\cmsorcid{0000-0002-7934-4038}}{10}
\AddAuthor{V.~Ivanchenko\cmsorcid{0000-0002-1844-5433}}{11}
\AddAuthor{M.~Janda\cmsorcid{0000-0001-5736-6183}}{1c}
\AddAuthor{A.~Karev}{8}
\AddAuthor{J.~Ka\v{s}par\cmsorcid{0000-0001-5639-2267}}{1b,8}
\AddAuthor{B.~Kaynak\cmsorcid{0000-0003-3857-2496}}{c}
\AddAuthor{J.~Kopal\cmsorcid{0000-0001-9751-7409}}{8}
\AddAuthor{V.~Kundr\'{a}t\cmsorcid{0000-0003-2868-5550}}{1b}
\AddAuthor{S.~Lami\cmsorcid{0000-0001-9492-0147}}{6a}
\AddAuthor{R.~Linhart}{1a}
\AddAuthor{C.~Lindsey}{10}
\AddAuthor{M.V.~Lokaj\'{i}\v{c}ek\cmsorcid{0000-0002-2052-1220}$^{\textrm{\dag,}}$}{1b}
\AddAuthor{L.~Losurdo\cmsorcid{0000-0002-4964-7951}}{6c}
\AddAuthor{F.~Lucas~Rodr\'{i}guez}{8}
\AddAuthor{M.~Macr\'{i}}{\dag,5a}
\AddAuthor{M.~Malawski\cmsorcid{0000-0001-6005-0243}}{7}
\AddAuthor{N.~Minafra\cmsorcid{0000-0003-4002-1888}}{10}
\AddAuthor{S.~Minutoli}{5a}
\AddAuthor{K.~Misan\cmsorcid{0009-0007-2416-042X}}{7}
\AddAuthor{T.~Naaranoja\cmsorcid{0000-0001-5797-7929}}{2a,2b}
\AddAuthor{F.~Nemes\cmsorcid{0000-0002-1451-6484}}{3a,3b,8}
\AddAuthor{H.~Niewiadomski}{9}
\AddAuthor{T.~Nov\'{a}k\cmsorcid{0000-0001-6253-4356}}{3b}
\AddAuthor{E.~Oliveri\cmsorcid{0000-0002-0832-6975}}{8}
\AddAuthor{F.~Oljemark\cmsorcid{0000-0003-0121-2761}}{2a,2b}
\AddAuthor{M.~Oriunno}{d}
\AddAuthor{K.~\"{O}sterberg\cmsorcid{0000-0003-4807-0414}}{2a,2b}
\AddAuthor{P.~Palazzi\cmsorcid{0000-0002-4861-391X}}{8}
\AddAuthor{V.~Passaro\cmsorcid{0000-0003-0802-4464}}{4a,4c}
\AddAuthor{Z.~Peroutka}{1a}
\AddAuthor{J.~Proch\'{a}zka\cmsorcid{0000-0002-2774-2245}}{1b}
\AddAuthor{M.~Quinto\cmsorcid{0000-0002-6363-6132}}{4a,4b}
\AddAuthor{E.~Radermacher}{8}
\AddAuthor{E.~Radicioni\cmsorcid{0000-0002-2231-1067}}{4a}
\AddAuthor{F.~Ravotti\cmsorcid{0000-0002-5709-6934}}{8}
\AddAuthor{C.~Royon\cmsorcid{0000-0002-7672-9709}}{10}
\AddAuthor{G.~Ruggiero}{8}
\AddAuthor{H.~Saarikko}{2a,2b}
\AddAuthor{V.D.~Samoylenko}{e}
\AddAuthor{A.~Scribano\cmsorcid{0000-0002-4338-6332}}{6a,10}
\AddAuthor{J.~\v{S}irok\'{y}}{1a}
\AddAuthor{J.~Smajek}{8}
\AddAuthor{W.~Snoeys\cmsorcid{0000-0003-3541-9066}}{8}
\AddAuthor{R.~Stefanovitch}{8}
\AddAuthor{J.~Sziklai}{3a}
\AddAuthor{C.~Taylor\cmsorcid{0000-0001-6816-8051}}{9}
\AddAuthor{E.~Tcherniaev\cmsorcid{0000-0002-3685-0635}}{11}
\AddAuthor{N.~Turini\cmsorcid{0000-0002-9395-5230}}{6c}
\AddAuthor{O.~Urban}{1a}
\AddAuthor{V.~Vacek\cmsorcid{0000-0001-9584-0392}}{1c}
\AddAuthor{O.~Vavroch}{1a}
\AddAuthor{J.~Welti}{2a,2b}
\AddAuthor{J.~Williams\cmsorcid{0000-0002-9810-7097}}{10}
\AddAuthorLast{J.~Zich}{1a}
\vskip 4pt plus 4pt
\let\thefootnote\relax
\newcommand{\AddInstitute}[2]{${}^{#1}$#2\\}
\newcommand{\AddExternalInstitute}[2]{\footnote{${}^{#1}$ #2}}
\noindent
\dag Deceased\\
\AddInstitute{1a}{University of West Bohemia, Pilsen, Czech Republic}
\AddInstitute{1b}{Institute of Physics of the Academy of Sciences of the Czech Republic, Prague, Czech Republic}
\AddInstitute{1c}{Czech Technical University, Prague, Czech Republic}
\AddInstitute{2a}{Helsinki Institute of Physics, University of Helsinki, Helsinki, Finland}
\AddInstitute{2b}{Department of Physics, University of Helsinki, Helsinki, Finland}
\AddInstitute{3a}{Wigner Research Centre for Physics, RMKI, Budapest, Hungary}
\AddInstitute{3b}{MATE Institute of Technology KRC, Gy\"{o}ngy\"{o}s, Hungary}
\AddInstitute{4a}{INFN Sezione di Bari, Bari, Italy}
\AddInstitute{4b}{Dipartimento Interateneo di Fisica di Bari, University of Bari, Bari, Italy}
\AddInstitute{4c}{Dipartimento di Ingegneria Elettrica e dell'Informazione --- Politecnico di Bari, Bari, Italy}
\AddInstitute{5a}{INFN Sezione di Genova, Genova, Italy}
\AddInstitute{5b}{Universit\`{a} degli Studi di Genova, Genova, Italy}
\AddInstitute{6a}{INFN Sezione di Pisa, Pisa, Italy}
\AddInstitute{6c}{Universit\`{a} degli Studi di Siena and Gruppo Collegato INFN di Siena, Siena, Italy}
\AddInstitute{7}{Akademia G\'{o}rniczo-Hutnicza (AGH) University of Science and Technology, Krakow, Poland}
\AddInstitute{8}{CERN, Geneva, Switzerland}
\AddInstitute{9}{Case Western Reserve University, Department of Physics, Cleveland, Ohio, USA}
\AddInstitute{10}{The University of Kansas, Lawrence, Kansas, USA}
\AddInstitute{11}{Authors affiliated with an institute or an international laboratory covered by a cooperation agreement with CERN}
\AddExternalInstitute{a}{INRNE-BAS, Institute for Nuclear Research and Nuclear Energy, Bulgarian Academy of Sciences, Sofia, Bulgaria}
\AddExternalInstitute{b}{Department of Atomic Physics, E\"{o}tv\"{o}s Lor\'{a}nd University, Budapest, Hungary}
\AddExternalInstitute{c}{Istanbul University, Istanbul, Turkey}
\AddExternalInstitute{d}{SLAC, Stanford University, California, USA}
\AddExternalInstitute{e}{Authors affiliated with an institute or an international laboratory covered by a cooperation agreement with CERN}\end{sloppypar}
%%% END EDITABLE REGION %%%
% skeleton_end
\end{document}